\documentclass{article}

\usepackage{arxiv}
\usepackage{array}
\usepackage[utf8]{inputenc} 
\usepackage[T1]{fontenc}    
\usepackage{hyperref}       
\usepackage{url}            
\usepackage{booktabs}       
\usepackage{amsfonts}       
\usepackage{nicefrac}       
\usepackage{microtype}      
\usepackage{graphicx}
\graphicspath{ {./fig/} }
\usepackage[square,numbers]{natbib}
\usepackage{multirow}
\usepackage{doi}
\usepackage{nomencl}
\usepackage[ruled]{algorithm2e}
\usepackage{float}
\usepackage[table,xcdraw]{xcolor} 
\usepackage{tcolorbox}
\usepackage{enumitem}
\usepackage{amsmath}

\title{Large Language Models for Combinatorial Optimization of Design Structure Matrix}

\date{Current Version: Apr 1, 2026. \thanks{This is an extended version prepared for journal submission. \href{https://doi.org/10.1017/pds.2025.10234}{An earlier version} of this work was presented at the 25th International Conference on Engineering Design (ICED 25), held in Dallas, Texas, USA.}}
\author{
{
\hspace{1mm}Shuo Jiang}\thanks{Comments are welcome: \texttt{shuojiangcn@gmail.com}} \\
	Department of Systems Engineering \\
        City University of Hong Kong, Hong Kong \\
	\texttt{shuo.jiang@cityu.edu.hk} \\
	\And
	{\hspace{1mm}Min Xie} \\
        Department of Systems Engineering \\
        City University of Hong Kong, Hong Kong \\
	\texttt{xiemin@cityu.edu.hk} \\
        \And
	{\hspace{1mm}Jianxi Luo$^{*}$} \\
        Department of Systems Engineering \\
        City University of Hong Kong, Hong Kong \\
	\texttt{jianxi.luo@cityu.edu.hk} \\
}

\renewcommand{\headeright}{Large Language Models for Combinatorial Optimization of Design Structure Matrix}
\renewcommand{\undertitle}{}
\renewcommand{\shorttitle}{}

\begin{document}
\maketitle

\begin{abstract}
	In complex engineering systems, the dependencies among components or development activities are often modeled and analyzed using Design Structure Matrix (DSM). Reorganizing elements within a DSM to minimize feedback loops, thereby enhancing modularity or process efficiency, constitutes a challenging Combinatorial Optimization (CO) problem in engineering design and operations. As problem sizes increase and dependency networks become more intricate, traditional optimization methods that rely solely on mathematical heuristics often fail to capture the contextual nuances and struggle to deliver effective solutions. In this study, we explore the potential of Large Language Models (LLMs) to address such CO problems by leveraging their capabilities for advanced reasoning and contextual understanding. We propose a novel LLM-based framework that integrates network topology with contextual domain knowledge for iterative optimization of DSM sequencing, a representative CO problem in this domain. Experiments on various DSM cases demonstrate that our proposed method consistently achieves faster convergence and superior solution quality compared to both stochastic and deterministic baselines. Notably, incorporating contextual domain knowledge significantly enhances optimization performance regardless of the chosen LLM backbone. This study demonstrates the potential of LLMs as a promising foundation for advancing knowledge-informed optimization in engineering design.
\end{abstract}

\keywords{Large Language Models \and Artificial Intelligence \and Design Structure Matrix \and Combinatorial Optimization \and Engineering Design \and Systems Engineering}

\section{Introduction}
\label{sec1}

Design Structure Matrix (DSM) is a well-established modeling tool in engineering design \cite{Eppinger2012,Steward1981}. It encodes relationships among elements, such as activities, components, or parameters, in a compact matrix \cite{Steward1981}. Reordering the sequence of nodes in a DSM to minimize feedback loops and improve modularity can improve system performance, reduce development risk, and increase design efficiency \cite{Choi2011,Eppinger2012}. Consider a simple DSM with three tasks: A, B, and C, where B depends on outputs from A, and C depends on outputs from B. In the sequence [A, B, C], all dependencies flow forward with no backward links. In contrast, the sequence [B, A, C] places B before A despite B requiring A's output, creating a backward dependency that necessitates rework and reduces process efficiency.

Figure 1 illustrates a real-world example of a design activity DSM before and after sequencing \cite{Amen1999}. DSM sequencing is a typical combinatorial optimization (CO) problem, and also an NP-hard problem. As matrix sizes and dependency intricacy increase, traditional optimization methods that rely on heuristics and rule-based algorithms often struggle to find effective solutions \cite{Eppinger1994,Qian2011}.

\begin{figure}[H]
	\centering
	\includegraphics[width=16.5cm]{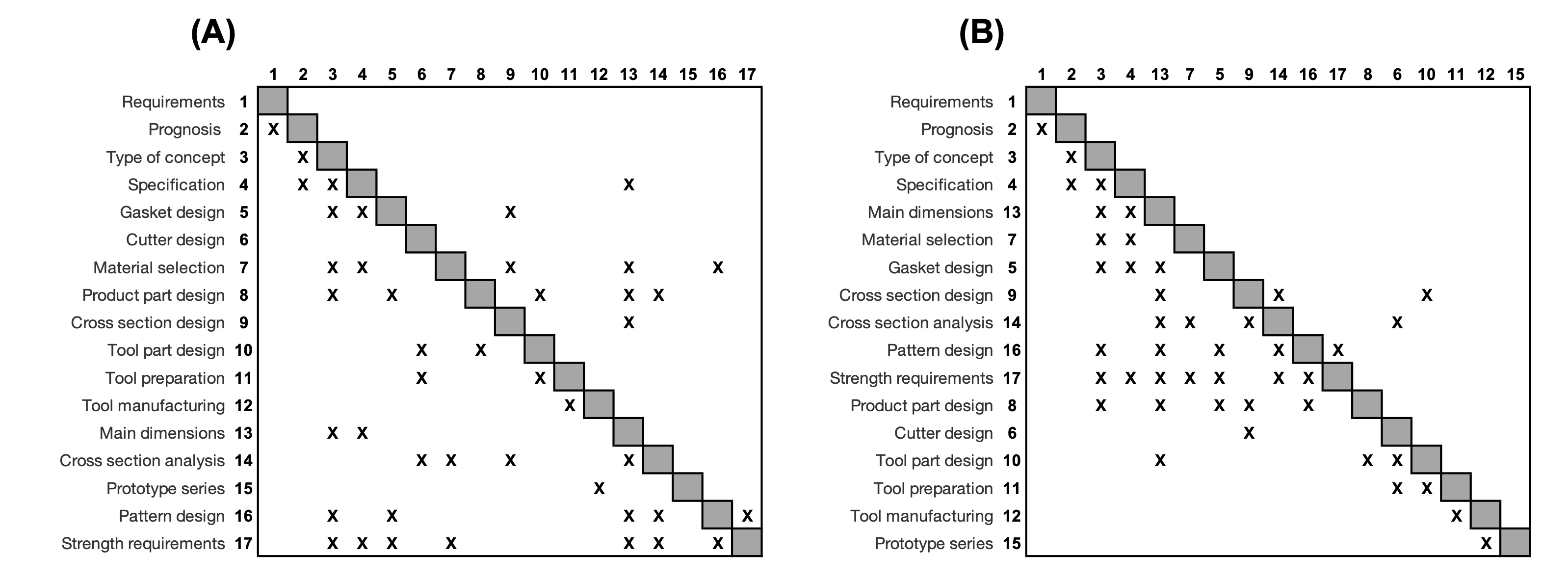}
	\caption{Illustration of a Design Activity DSM: (A) Pre-Sequencing; (B) Post-Sequencing}
	\label{fig:fig1}
\end{figure}

Recent advancements in Large Language Models (LLMs) have demonstrated their powerful capabilities in natural language generation, semantic understanding, instruction following, and complex reasoning \cite{Chang2024,Wei2022}. Furthermore, studies have shown that LLMs can be used for continuous and concrete optimization \cite{Liu2024eoh,Liu2024llmaseo,Romera-Paredes2024,Yang2024}. For instance, researchers from Google DeepMind utilized LLMs as optimizers and evaluated their effectiveness on classic CO problems, such as Traveling Salesman Problems \cite{Yang2024}. In the field of engineering, researchers have also applied generative artificial intelligence (GenAI) techniques to address engineering optimization problems \cite{Picard2024}. In addition, previous studies have shown that LLMs possess extensive engineering domain knowledge learned during pretraining on diverse data resources, which enhances their applicability in engineering contexts \cite{Jiang2025,Makatura2024,Mei2024,Zhu2023}.

Therefore, the ability of LLMs to combine mathematical and semantic reasoning, along with their inherent extensive domain knowledge, motivated us to explore their potential for DSM optimization tasks in engineering design, where contextual information is useful for identifying high-quality solutions. Our hypotheses are: (1) LLMs can be effectively applied to DSM-based combinatorial optimization tasks, and (2) incorporating contextual domain knowledge can further enhance the proposed LLM-based method by supporting mathematical reasoning with semantic insights.

This paradigm, which leverages both semantic and mathematical reasoning, introduces a novel approach to combinatorial optimization beyond the reach of traditional purely mathematical methods. To validate the proposed method, we conduct experiments on various engineering-related DSM cases collected from peer-reviewed literature. The experimental results demonstrate the practical applicability of our proposed LLM-based approach for combinatorial optimization of DSM. The main contributions of this paper are summarized as follows:

\begin{enumerate}[label=(\arabic*)]
  \item We propose a novel LLM-based optimization framework that integrates network topology and contextual domain knowledge for DSM sequencing optimization.
  \item We perform a comprehensive benchmarking analysis against both stochastic and deterministic algorithms, and systematically evaluate the impact of different backbone LLMs, validating the robustness and adaptability of our method across various engineering DSM cases.
  \item We show that incorporating domain knowledge significantly improves both solution quality and convergence speed in our method. To the best of our knowledge, this is the first study to integrate LLMs and DSM for engineering optimization.
\end{enumerate}

This paper is organized as follows. Section 2 reviews related work on DSM and LLMs for optimization. Section 3 presents our proposed LLM-based framework in detail. Section 4 describes the data, experiments, and benchmark methods. Section 5 presents the experimental results and discusses key insights and practical implications for engineering design. Section 6 summarizes the limitations of the study and discusses potential directions for future work. Finally, Section 7 concludes the paper.

\section{Background and Related Work}
\label{sec2}
\subsection{Design Structure Matrix for Engineering Design}

The DSM is a powerful tool that models the interactions and dependencies among elements within complex engineering systems \cite{Eppinger2012}. Originating from Steward’s pioneering work \cite{Steward1981}, DSMs provide a systematic framework to represent and manage dependencies among elements in complex systems, using a compact matrix format. Each row and column in the matrix correspond to a specific component, activity, or parameter, while the matrix entries illustrate the state and strength of their dependencies. DSM methodologies have evolved significantly, with researchers proposing advanced methods for managing complex engineering design processes, improving modularity, and enhancing overall system efficiency \cite{Eppinger1994}. Since then, DSMs have seen significant development in both academia and industry. DSMs have been successfully applied across many disciplines for engineering design, such as automotive \cite{Sequeira1991,Smith1997}, semiconductor \cite{Osborne1993}, aerospace \cite{Clarkson2000}, factory equipment \cite{Hameri1999}, manufacturing systems \cite{Luo2012,luo2019hierarchy}, and more \cite{Karniel2009,Ogura2019,Pasqual2012,Wong2024,Zhang2026}.

DSM can be categorized primarily into four types: component-based, team-based, activity-based, and parameter-based \cite{Browning2001}. The first two types are static DSMs, whereas the other two are time-based DSMs. Each category addresses specific aspects of engineering design, ranging from structuring task flows and team interactions to the relationships between technical parameters of system components. Static DSMs are typically analyzed using clustering algorithms, which identify modular groupings to minimize dependencies and facilitate parallel development. Time-based DSMs are usually analyzed using sequencing algorithms to streamline processes, reduce iterative feedback loops, and optimize the logical progression of activities.

The optimization of DSMs in engineering design is a typical CO problem, requiring efficient and effective methods to obtain high-quality solutions. Researchers have explored various algorithms for DSM applications, including heuristic, exact and hybrid approaches \cite{Ahmadi2001,Attari-Shendi2019,Kusiak1993,Lin2012,Qian2011,Rogers1996}. For example, Meier et al. \cite{Meier2006} proposed the Ordering messy Genetic Algorithm (OmeGA), which integrates modified GA and local search to enhance DSM sequencing. Qian et al. \cite{Qian2011} presented a hybrid heuristic-exact method based on block folding, which shows progressive improvement and performs efficiently on the DSM sequencing problem. These previous works have continuously advanced mathematical methods to enhance the performance and scalability of DSM optimization. However, while DSMs represent real-world engineering problems in a compact and abstract form, the underlying systems they model are embedded in rich contextual domain knowledge that can be valuable for guiding optimization. With the capabilities of large-scale foundation models, such information can be effectively incorporated into the optimization process. This paper contributes to this field by leveraging LLMs for DSM optimization, considering both semantic and mathematical reasoning.

\subsection{Large Language Models for Combinatorial Optimization}

Combinatorial optimization problems are frequently encountered across diverse engineering domains, where finding an optimal solution from a finite set often drives improvements in efficiency, cost, and performance \cite{Korte2011}. For instance, applications such as DNA barcoding and DNA assembly in synthetic biology \cite{Naseri2020}, as well as job scheduling in manufacturing \cite{Xidias2019}, rely heavily on effective solutions of CO problems. However, due to their NP-hard nature, these problems present substantial challenges, especially as complexity increases with larger problem sizes and more intricate dependencies. Traditionally, real-world CO problems are usually approached through the following process: the problem is first modeled formally and mathematically, then solved using specific optimization algorithms or heuristics, and finally interpreted within the context of practical engineering \cite{Pinedo2012}. This separation between problem-solving and interpretation stages limits the ability to capture the contextual nuances that are often essential for optimizing real-world problems.

Recent advancements have significantly enhanced the capabilities of LLMs \cite{Chang2024,Wei2022,JiangID2025}. LLMs acquire extensive common-sense and domain-specific knowledge during their pretraining on large-scale datasets, including publicly available code repositories, natural language corpora, and other multimodal data. This broad foundation is further refined through fine-tuning on various downstream tasks, enabling LLMs to adapt to specific applications. State-of-the-art LLMs such as Claude, GPT, Gemini, and DeepSeek models have demonstrated remarkable performance in understanding natural language, complex reasoning, and generalizing across a wide range of tasks \cite{anthropic2024claude,Liu2024deepseek,meta2024llama,openai2024gpt4turbo}.

LLMs have recently been leveraged to address CO problems \cite{Liu2024eoh,Liu2024llmaseo,Romera-Paredes2024,Yang2024}. Early explorations utilized prompt engineering to guide LLMs in solving classical CO problems, such as the traveling salesman problem and job-shop scheduling, demonstrating the feasibility of using LLMs for direct solution generation \cite{Liu2024llmaseo,Yang2024}. Later, inspired by automatic heuristic generation, researchers proposed hybrid approaches that combine LLMs with evolutionary computation to generate, adapt, and refine optimization heuristics \cite{Liu2024eoh,Romera-Paredes2024}. Compared with traditional optimization algorithms, these approaches benefit from LLMs’ flexible reasoning and knowledge-driven heuristic generation, enabling generalization across problem variants without handcrafted solvers.

However, most of these studies simplify CO problems into a formal representation, overlooking contextual domain-specific knowledge that is critical in real-world applications. Recent studies have demonstrated that LLMs can generate interpretable DSM structures directly from textual descriptions of engineering processes \cite{Koh2024}, and further work has shown that the quality of such retrieval is sensitive to factors such as writing style and entity naming conventions \cite{Koh2026}, highlighting the nuanced ways in which LLMs engage with domain-specific knowledge. These converging capabilities position LLMs as a promising yet underexplored tool for CO problems in engineering design, motivating the framework proposed in this paper \cite{Bordas2024}.

\section{Methodology}
\label{sec3}

In this section, we present our proposed LLM-based approach in detail and apply it to DSM sequencing, which aims to reduce backward dependencies and enhance system modularity in engineering workflows. Overall, the approach is designed to harness the generative and reasoning capabilities of LLMs, combining them with domain-specific contextual knowledge and a formal evaluation mechanism to iteratively drive the search toward high-quality solutions.

The process begins with the initialization of a valid solution (a complete ordering of all nodes in the DSM) that is randomly sampled from the total solution space. Each solution is then evaluated using a predefined evaluator that quantifies its quality based on specific criteria. Guided by this evaluation, the approach iteratively updates the solution base by leveraging the in-context learning capabilities of LLMs. Prompts of each round are crafted using both the network’s mathematical representation and contextual domain knowledge in natural language descriptions, along with meta-instructions and selected historical solutions. In practice, the DSM data used in this process are constructed from documented system architectures and task dependencies, which can be extracted from prior studies, design specifications, or expert elicitation. In each iteration, the LLM generates a new candidate solution, which is then validated and evaluated. Each validated solution and its corresponding score are recorded in the solution base for use in subsequent iterations. Upon meeting the termination condition, the highest-scoring solution in the solution base is returned as the final output. The overall workflow is illustrated in Figure 2.

\begin{figure}[H]
	\centering
	\includegraphics[width=16.5cm]{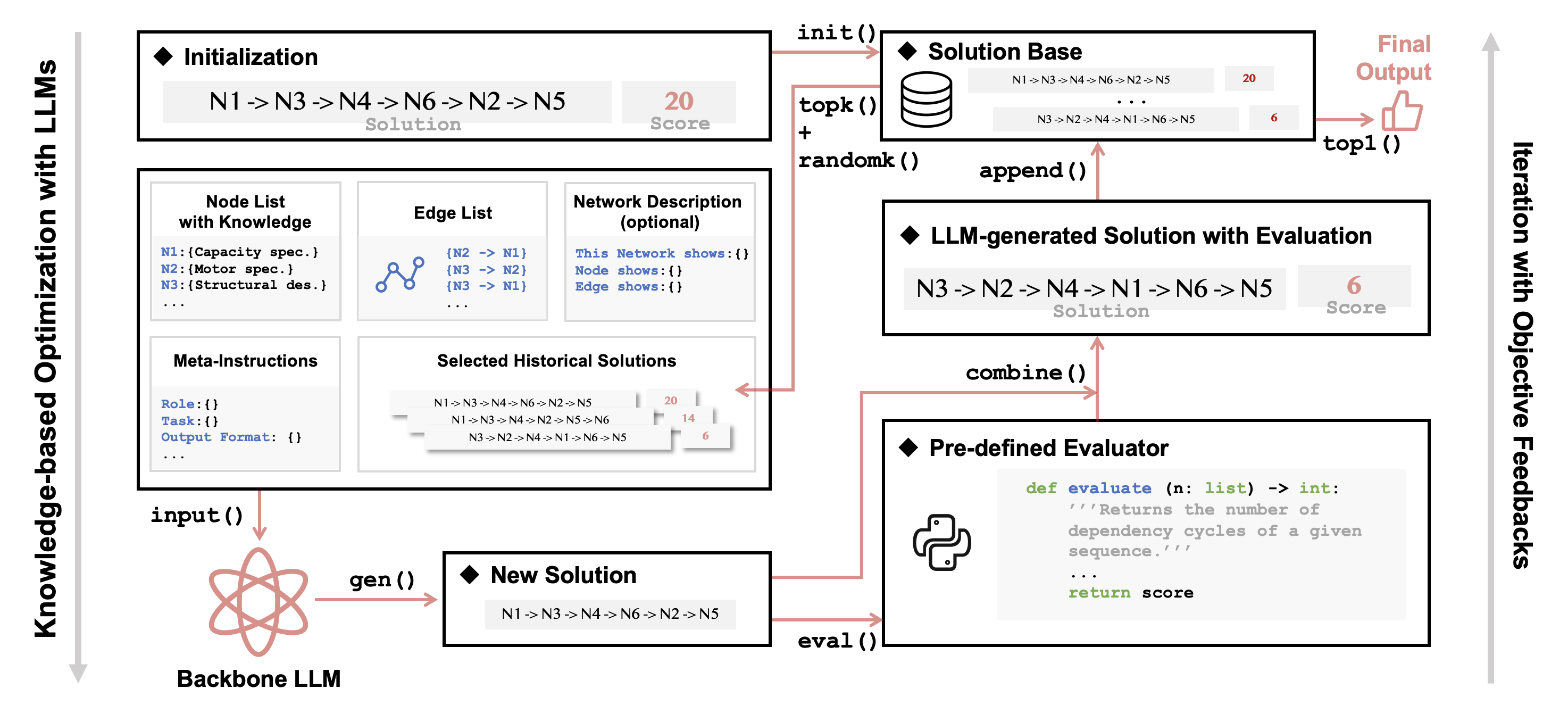}
	\caption{Overview of the LLM-based DSM Optimization Framework}
	\label{fig:fig2}
\end{figure}

\subsection{Initialization and Solution Sampling}

The initialization begins by randomly sampling a valid solution from the entire solution space. In the context of DSM sequencing, each solution corresponds to a complete and non-repetitive permutation of nodes. For instance, in a DSM with nodes {A, B, C}, one possible candidate sequence is [B, A, C]. This initial solution is evaluated using the predefined evaluator and added to the solution base. The solution base is a key component of the approach, serving three primary purposes: (1) storing all previously explored solutions along with their evaluation results ([\textit{Solution, Score}] pairs), (2) providing historical solutions for few-shot learning to guide the LLM’s generation in subsequent iterations, and (3) returning the top-performing solution as the final output at the end of the optimization process.

During subsequent iterations, a sampling strategy is employed to construct a candidate solution set for LLM in-context learning. Specifically, we select the top $K_p$ highest-scoring solutions and randomly sample additional $K_q$ solutions from the remaining $K_n-K_p$ entries in the solution base, where $K_n$ denotes the total number of stored solutions. The combined set of $K_p+K_q$ solutions is then used to construct prompts that guide the next round of LLM-based generation. The values of $K_p$ and $K_q$ are tunable hyperparameters that balance exploitation and exploration in the search process. These historical solutions are formatted as a list of [\textit{Solution, Score}] pairs in the prompt.

\begin{figure}
	\centering
	\includegraphics[width=16.5cm]{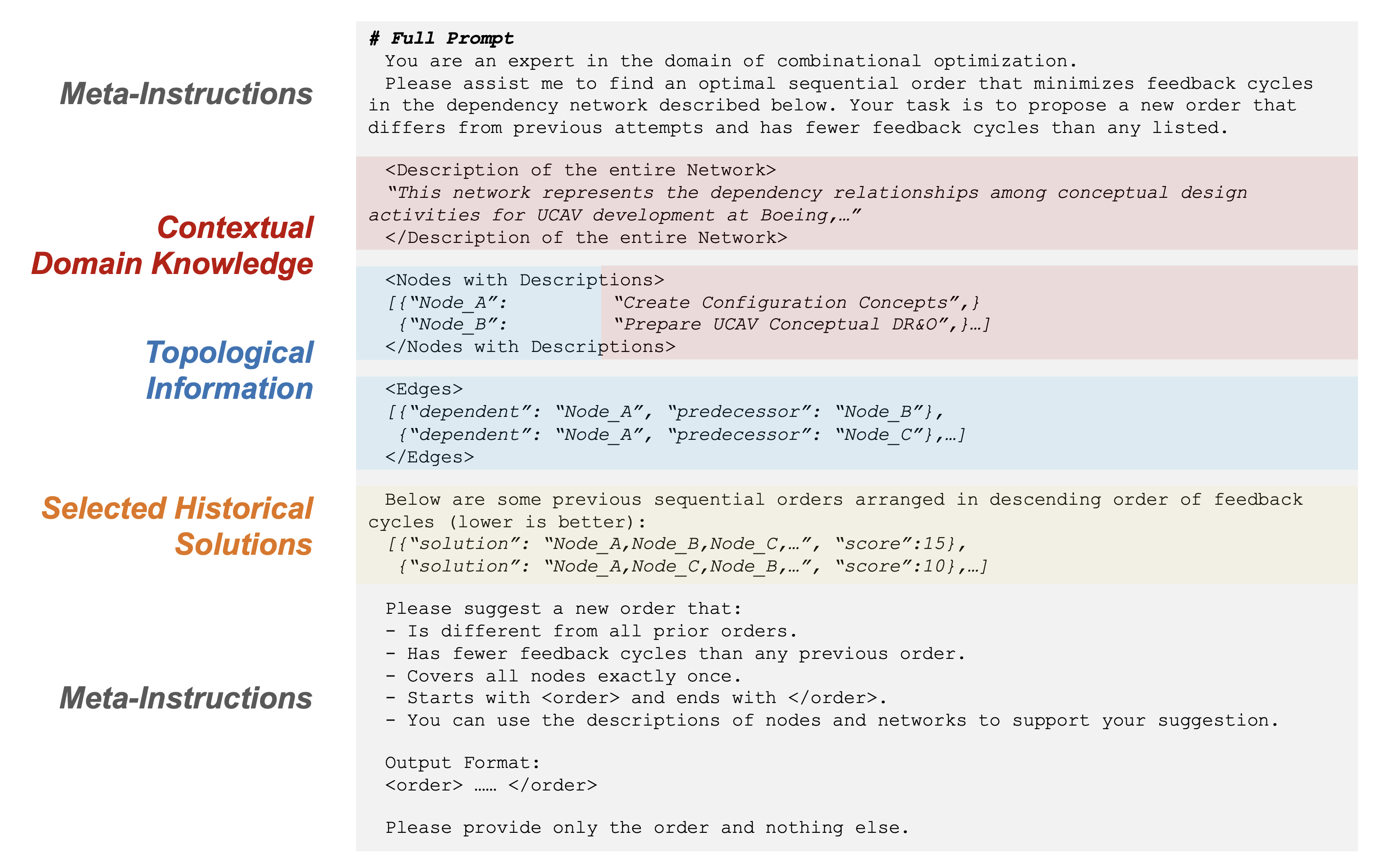}
	\caption{Structure of the input to LLMs for DSM sequencing optimization}
	\label{fig:fig3}
\end{figure}

\subsection{LLM-driven Generative Optimization}

In each iteration of the optimization process, the backend LLM is prompted with a structured input that incorporates both mathematical representations of the DSM and contextual knowledge related to the given DSM. The prompt includes the following four key components:

\textbf{(i) Topological Information:} This provides a mathematical description of the DSM. While multiple representations can be used to describe a complex network (e.g., an adjacency matrix, a dependency list, or an edge list), we adopt the edge list format for its simplicity and compatibility with sequence-based reasoning. To mitigate any ordering bias, all edges are randomly shuffled before being included in the prompt.

\textbf{(ii) Contextual Domain Knowledge:} This component includes the semantic information associated with each node (e.g., activity names, component functions) and an overall textual description of the network. These natural language elements convey the engineering context behind the DSM structure, enabling the LLM to reason beyond abstract graph features. For example, in an activity-based DSM, nodes represent distinct phases of the engineering design process.

\textbf{(iii) Meta-instructions:} To guide the generation behavior of the LLM, we adopt several frequently used prompt engineering strategies \cite{Zhao2023}, including role-playing, task specification, and output format specification. These instructions help structure the LLM’s responses and ensure syntactic and semantic consistency.

\textbf{(iv) Selected Historical Solutions:} As described in the previous section, a curated set of $K_p+K_q$ solutions are then sampled from the solution base and presented to the LLM for few-shot learning. This allows the model to generalize from prior high-quality and diverse solutions.

Upon receiving the complete input, the LLM generates a new candidate solution, which is then validated by a checker to ensure that it includes all nodes exactly once and adheres to the required sequencing constraints. Valid solutions are then evaluated using the predefined scoring function and subsequently appended to the solution base for use in future iterations.

Figure 3 illustrates an example prompt, where each component is explicitly represented: meta-instructions define the task and output requirements; contextual domain knowledge specifies node descriptions and system context; topological information provides the DSM’s node and edge list; and selected historical solutions offer prior orderings with scores. More details refer to Appendix 1.

\subsection{Evaluation of Generated Solutions}

Each candidate solution generated by the LLM is evaluated immediately upon creation using a predefined scoring function. In general, a wide range of objectives have been proposed for DSM sequencing in prior research, depending on whether the emphasis is on reducing iteration, improving concurrency, lowering cost, or enhancing modularity. Examples include minimizing feedbacks or weighted feedbacks \cite{kusiak1993decomposition,Steward1981}, minimizing feedback length \cite{Gebala1991}, and more \cite{Altus1996,Baldwin2000}. In this study, we adopt the most widely used objective: minimizing the number of feedback loops (i.e., non-zero entries above the diagonal of the DSM adjacency matrix). This criterion has been consistently employed in the literature as a baseline indicator of process rework, making it appropriate for evaluating the feasibility of an LLM-driven optimization approach.

The resulting score is then used to rank solutions in the solution base and to guide sampling and prompt construction in subsequent iterations. For optimization problems other than DSM sequencing, the evaluation criteria can be adapted based on the structural characteristics and design objectives of the specific problem. For instance, in modularity optimization (i.e., the partitioning task), the metric may focus on maximizing intra-module cohesion while minimizing inter-module dependencies. When the predefined number of iterations is reached or the evaluation score meets a known optimal threshold, the iteration process is terminated. The best-performing solution is then returned as the final output.

\section{Experiments}
\label{sec4}
\subsection{Data}

We collected four DSM cases for experiments, which are categorized into two types following established classifications \cite{Browning2001}: (1) Activity-based DSMs, which represent the input-output relationships among tasks or activities within a project; and (2) Parameter-based DSMs, which illustrate the precedence and dependencies among design parameters.

For activity-based DSMs, we collected two cases:

\begin{itemize}
    \item The Unmanned Combat Aerial Vehicle (UCAV) DSM includes 12 conceptual design activities conducted at Boeing \cite{Browning1998}.
    \item The Microfilm Cartridge DSM was derived from Kodak’s Cheetah project and consists of 13 major development tasks \cite{Eppinger2016}.
\end{itemize}

Both DSMs were constructed through structured interviews with domain experts and were subsequently reviewed and validated to ensure accuracy and consistency.

For parameter-based DSMs, we collected two cases:

\begin{itemize}
    \item The Heat Exchanger DSM, which includes 17 components related to core thermal exchange functions \cite{Amen1999}.
    \item The Automobile Brake System DSM, which comprises 14 design parameters reflecting braking mechanisms and their dependencies \cite{Black1990}.
\end{itemize}

Both DSMs were developed by first identifying key product components, followed by interviews with the respective design engineers to define parameters and establish directional relationships.

From the original references, we extracted three types of information for each DSM: (1) the names of all nodes (activities or parameters), (2) the edge list derived from the adjacency matrix, and (3) a natural language description summarizing the system-level context of each network. The data were maintained in their original structure to ensure fidelity and reproducibility. Data formatting details are provided in Appendix 1.

The characteristics of the four DSMs are summarized in Table 1. Overall, the number of nodes $N$ ranges from 12 to 17, and the number of edges $E$ varies between 32 and 47. We also report network-level metrics including diameter, density, average degree, clustering coefficient, and average path length to present topological complexity. For example, the UCAV DSM exhibits a high network density of 0.712 and a clustering coefficient of 0.773, reflecting tightly coupled design tasks. In contrast, the Heat Exchanger DSM is more sparsely connected, with clearer modular separations among design parameters.

\begin{table}[H]
    \caption{Characteristics of Four DSMs}
    \centering
    \renewcommand{\arraystretch}{1.5}
    \begin{tabular}{m{4.3cm}<{\arraybackslash} m{0.55cm}<{\centering\arraybackslash} m{0.55cm}<{\centering\arraybackslash} m{1.45cm}<{\centering\arraybackslash} m{1.3cm}<{\centering\arraybackslash} m{1.3cm}<{\centering\arraybackslash} m{1.7cm}<{\centering\arraybackslash} m{1.95cm}<{\centering\arraybackslash}}
        \toprule
        \textbf{} & \textbf{N} & \textbf{E} & \textbf{Network Diameter} & \textbf{Network Density} & \textbf{Average Degree} & \textbf{Clustering Coefficient} & \textbf{Average Path Length} \\
        \midrule

        \multicolumn{8}{l}{\textbf{Activity-Based DSMs}} \\
        Unmanned Aerial Vehicle \cite{Browning1998} & 12 & 47 & 2 & 0.712 & 7.833 & 0.773 & 1.288 \\
        Microfilm Cartridge \cite{Eppinger2016}     & 13 & 41 & 3 & 0.526 & 6.308 & 0.682 & 1.577 \\

        \midrule

        \multicolumn{8}{l}{\textbf{Parameter-Based DSMs}} \\
        Heat Exchanger \cite{Amen1999}              & 17 & 41 & 7 & 0.302 & 4.824 & 0.457 & 2.397 \\
        Automobile Brake System \cite{Black1990}    & 14 & 32 & 4 & 0.352 & 4.571 & 0.414 & 1.824 \\

        \bottomrule
    \end{tabular}
    \label{tab:table1}
\end{table}

\subsection{Evaluation Metrics}

DSM sequencing can be assessed under multiple objectives depending on the design context. Prior studies have considered criteria such as reducing iteration, enhancing concurrency, minimizing development time and cost, or improving modularity of the system \cite{Baldwin2000,Gebala1991,kusiak1993decomposition,Steward1981}. In this study, we adopt the minimization of feedback loops as the evaluation metric, as it is the widely adopted measure in the DSM literature \cite{Steward1981}. It is noteworthy that the sequence with the fewest feedbacks is not always the practically optimal one, but it serves as a widely adopted and reasonable foundation for evaluating our approach. To explore a more nuanced objective, we also evaluate our framework using a weighted feedback metric in Section 5.4, which accounts for both the presence and positional span of each feedback dependency.

To evaluate a given sequence, we first construct an $n \times n$ asymmetric adjacency matrix $A$ based on the edge list of the network, where $n$ denotes the number of nodes. Each binary entry $a_{ij}$ in the matrix indicates whether node $i$ depends on node $j$, with $a_{ij} = 1$ representing a directed edge from $j$ to $i$. The sequencing task involves reordering the nodes such that the number of non-zero entries above the main diagonal in $A$ is minimized. These entries correspond to feedback dependencies that disrupt ideal information or task flow. Formally, given a sequence $s$, the evaluation score is computed as:

$$
\text{Score}(s) = \sum_{i=1}^{n} \sum_{j=i+1}^{n} a_{ij}
$$

Lower values correspond to improved modularization or reduced rework potential in engineering workflows. Given the heuristic and iterative nature of our approach, the number of feedback loops is computed at each iteration for every valid solution. This allows us to track convergence and identify high-quality solutions progressively. For a fair comparison across methods, we also report the best achieved value within a fixed number of trials for each configuration in our experiments.

During performance benchmarking, we complement this primary metric with statistical reporting of mean and standard deviation over multiple trials. For convergence analysis, we record the number of unique sequences explored before reaching the optimal score, thereby capturing both efficiency and solution quality.

\subsection{Experiment Setup}

To ensure the robustness of results, each method was executed 10 times with different random seeds. In the solution sampling process, we set the top-performing selection parameter $K_p=5$ and the random sampling parameter $K_q=5$. Each node in the DSM was assigned a unique identifier, composed of five randomly generated alphanumeric characters, to ensure independence from domain-specific naming. The maximum number of iterations for all experiments was fixed at 20. \textit{Claude-Sonnet-3.5-20241022} was used as the main backbone LLM, and all model parameters were kept at their default settings \cite{anthropic2024claude}. This model was selected as it demonstrates the most consistent performance across different DSM types and trial budgets in our ablation study, making it a reliable default choice for the main experiments. In practice, users may select alternative backbone LLMs based on their specific deployment constraints or cost considerations, as discussed in Section 5.3.

To further evaluate the generalizability of the framework, we conducted additional experiments across three dimensions: a backbone LLM ablation study comparing eight models; a scalability test on a larger and denser DSM case; and an evaluation under an alternative optimization objective using a weighted feedback metric. Unless otherwise specified, all additional experiments follow the same protocol described above.

\subsection{Baseline Methods}

To benchmark the performance of our proposed approach, we compare it against both stochastic and deterministic optimization approaches commonly used for CO problems and DSM sequencing tasks.

\textbf{Stochastic Methods}. We adopt the classic Genetic Algorithm (GA), a widely used probabilistic optimization method \cite{Alhijawi2024}. Three configurations were tested to explore different optimization strategies:

\begin{enumerate}[label=(\arabic*)]
    \item Exploration-focused: Prioritizes diversity to avoid premature convergence.
    \item Exploitation-focused: Focuses on intensifying search around high-performing regions.
    \item Balanced setting: Seeks a compromise between exploration and exploitation.
\end{enumerate}

In addition, we include the Ordered Messy Genetic Algorithm (OmeGA), which extends the standard GA with purpose-specific enhancements for DSM sequencing \cite{Meier2006}. OmeGA first decomposes the DSM into strongly connected components, applies topological sorting for inter-block ordering, and then performs genetic optimization with integrated local search within each block. Detailed settings are described in Appendix 2.

\textbf{Deterministic Methods}. Deterministic baselines reorder nodes based on fixed structural metrics. Since the calculation of each measure is deterministic, the resulting solution is consistent. Five representative deterministic algorithms were selected:

\begin{enumerate}[label=(\arabic*)]
    \item \textbf{Out-In Degree} \cite{Crofts2011}: Orders nodes by the difference between out-degree and in-degree.
    \item \textbf{Eigenvector} \cite{Dietzenbacher1992}: Uses the Perron vector of the adjacency matrix to rank nodes.
    \item \textbf{Walk-based (Exponential)} \cite{Estrada2005}:Computes $F(\mathbf{A}) = \exp(\mathbf{A})$, capturing long-range connectivity.
    \item \textbf{Walk-based (Resolvent)} \cite{Estrada2010}: Computes $F(\mathbf{A}) = (\mathbf{I} - \delta\mathbf{A})^{-1}$, where $\delta=0.025$ represents the probability that a message will successfully traverse an edge. The value of $\delta$ is adopted based on prior work.
    \item \textbf{Visibility} \cite{MacCormack2012}: Computes $F(\mathbf{A}) = \sum_{k=0}^{n} \mathbf{A}^k$, aggregating all path-based dependencies, then binarizes the result.
\end{enumerate}

For deterministic methods (3), (4), and (5), nodes are ranked based on the row sums of F(A); ties are resolved using column sums.

\textbf{Ablation Setting}. To assess the role of contextual domain knowledge, we implement a variant of our method that removes all semantic information from the input prompt. In this version, LLMs are guided solely by the DSM’s topological information, without node names or network descriptions. This setup enables us to assess the independent contribution of domain knowledge to optimization performance.

\section{Results and Discussion}
\label{sec5}
\subsection{Performance Benchmarking Against Baseline Methods}

Using the number of feedback loops as the metric (described in Section 4.2), we compared the performance of our proposed LLM-based method against both stochastic and deterministic baselines across four DSM cases, as shown in Table 2. The entries represent the number of feedback loops. Note that fractional values arise because the results are averaged over multiple independent runs, and the ± terms indicate one standard deviation, reflecting variability in performance.

Our methods were evaluated in three configurations: single-trial, 5-trial, and 20-trial runs, each tested with and without the inclusion of contextual domain knowledge. Among deterministic baselines, which are inherently single-trial and produce consistent outputs, our single-trial method with knowledge consistently outperformed all baseline methods, except for the Automobile Brake System DSM, where the Visibility-based method achieved a lower score than our single-trial configuration. In that same case, our method achieved identical best results when run under the 5-trial and 20-trial configurations.
When compared to stochastic baselines, the 20-trial LLM-based method matched or outperformed all GA variants and OmeGA in three of the four cases. Although OmeGA reached the optimal solution in the Heat Exchanger DSM, our method delivered strong performance surpassing all GA variants despite being limited to only 20 trials. In the remaining three DSM cases, our LLM-based method consistently found the optimal solutions within 20 trials, demonstrating its strong effectiveness under constrained iteration budgets. With only five trials, our method delivered competitive results compared to all stochastic methods in most cases. These findings highlight the efficiency of the proposed approach in generating high-quality solutions with minimal search effort.

\begin{table}[H]
    \caption{Performance Comparison with Baseline Methods}
    \centering
    \renewcommand{\arraystretch}{1.3}
    \setlength{\tabcolsep}{6pt} 
    \begin{tabular}{l m{2.3cm}<{\centering} m{2.3cm}<{\centering} m{2.3cm}<{\centering} m{2.3cm}<{\centering}}
        \toprule
        \textbf{Methods} & \multicolumn{2}{c}{\textbf{Activity-Based DSMs}} & \multicolumn{2}{c}{\textbf{Parameter-Based DSMs}} \\
        \cmidrule(lr){2-3} \cmidrule(lr){4-5}
        & \textbf{Unmanned Aerial Vehicle} & \textbf{Microfilm Cartridge} & \textbf{Heat Exchanger} & \textbf{Automobile Brake System} \\
        \midrule
        \textbf{Stochastic Methods\footnotemark[1]} & & & & \\
        GA-Exploration@500 & 7.9±1.1 & 9.1±0.9 & 7.2±0.9 & 5.2±0.9 \\
        GA-Exploration@10K & \textbf{6.0±0.0} & 8.1±0.3 & 5.7±1.0 & 3.8±0.7 \\
        GA-Balance@500 & 8.8±1.3 & 10.9±1.0 & 8.9±1.7 & 6.8±1.6 \\
        GA-Balance@10K & \textbf{6.0±0.0} & 8.4±0.5 & 6.2±1.2 & 4.2±1.2 \\
        GA-Exploitation@500 & 14.1±2.9 & 16.7±2.1 & 12.5±2.2 & 11.4±2.5 \\
        GA-Exploitation@10K & 7.4±1.5 & 9.9±1.4 & 7.6±1.4 & 6.6±1.7 \\
        OmeGA@500 & 6.3±0.9 & \textbf{8.0±0.0} & 3.5±0.5 & \textbf{3.0±0.0} \\
        OmeGA@10K & \textbf{6.0±0.0} & \textbf{8.0±0.0} & \textbf{3.0±0.0} & \textbf{3.0±0.0} \\
        \midrule
        \textbf{Deterministic Methods\footnotemark[2]} & & & & \\
        Out-In Degree \cite{Crofts2011} & 10.0±0.0 & 12.3±0.5 & 10.3±0.6 & 5.7±0.9 \\
        Eigenvector \cite{Dietzenbacher1992} & 15.0±0.0 & 13.8±0.7 & 13.1±1.7 & 11.1±1.4 \\
        Walk-based (Exponential) \cite{Estrada2005} & 15.0±0.0 & 12.0±0.0 & 8.0±0.0 & 11.0±0.0 \\
        Walk-based (Resolvent) \cite{Estrada2010} & 9.0±0.0 & 12.0±0.0 & 8.0±0.0 & 11.0±0.0 \\
        Visibility \cite{MacCormack2012} & 25.6±2.2 & 8.8±0.7 & 6.0±1.3 & \textbf{3.0±0.0} \\
        \midrule
        \textbf{LLM-based Methods (Ours)} & & & & \\
        Single-trial with knowledge & 6.6±0.7 & \textbf{8.0±0.0} & 4.8±0.6 & 4.4±0.9 \\
        Single-trial without knowledge & 11.9±3.4 & 8.1±0.3 & 5.8±1.1 & 5.9±1.4 \\
        5-trial with knowledge & 6.1±0.3 & \textbf{8.0±0.0} & 4.0±0.4 & \textbf{3.0±0.0} \\
        5-trial without knowledge & 7.5±0.9 & \textbf{8.0±0.0} & 4.9±0.5 & 4.0±1.0 \\
        20-trial with knowledge & \textbf{6.0±0.0} & \textbf{8.0±0.0} & \textbf{3.6±0.5} & \textbf{3.0±0.0} \\
        20-trial without knowledge & 6.4±0.7 & \textbf{8.0±0.0} & 4.1±0.3 & 3.4±0.7 \\
        \bottomrule
    \end{tabular}
    \label{tab:table2}
\end{table}
\footnotetext[1]{For stochastic methods, GA@500 denotes the performance after 500 solutions have been explored (evaluated), and GA@10K denotes the performance after 10,000 solutions, consistent with Figure 4. Other GA variants are defined similarly.}
\footnotetext[2]{All deterministic methods can be regarded as single-trial approaches, as they produce consistent outcomes for identical inputs across repeated executions. However, certain nodes might share the same measures. For these nodes, a random order is applied, and the final statistical evaluation results are obtained from 10 runs.}

\begin{figure}[H]
	\centering
	\includegraphics[width=16.5cm]{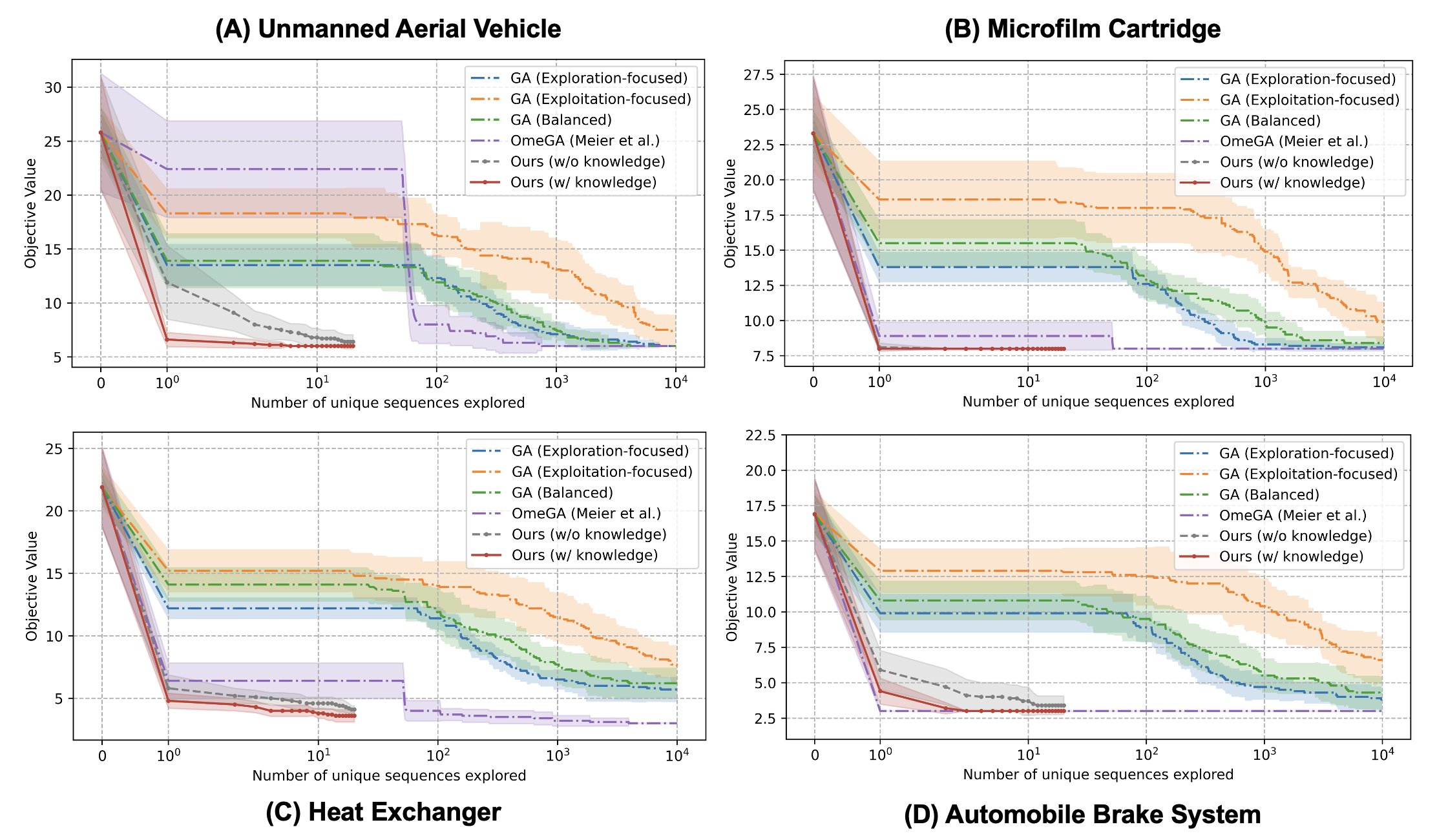}
	\caption{Convergence Behavior of LLM-based Methods and Benchmark Approaches}
	\label{fig:fig4}
\end{figure}

\subsection{Convergence Behavior and Search Efficiency}

We evaluated the convergence of our LLM-based methods in comparison with three variants of GA and the OmeGA algorithm, as shown in Figure 4. Because of their differing operational mechanisms, each GA iteration evaluates a set of candidate solutions generated through stochastic variation operators. Specifically, each GA generation explores 50 unique solutions, corresponding to its population size (i.e., the number of candidate solutions maintained per generation). Similarly, each generation of the OmeGA algorithm evaluates 50 unique solutions per block, followed by additional evaluations during the local search phase for the best individual in each generation. In contrast, our LLM-based approach produces one unique solution per iteration without population-based search. Although the maximum number of generations for both GA and OmeGA was set to 2,000 (refer to Appendix 2), the convergence comparison in Figure 4 focuses on the first 10,000 unique solutions to ensure clarity of visualization. The shaded bands around each curve in Figure 4 indicate one standard deviation across multiple independent runs.

Across all four DSM cases, our methods consistently demonstrate superior convergence speed relative to all GA variants and remain highly competitive with the purpose-specific OmeGA algorithm. The LLM-based methods reach substantially lower objective values with far fewer explored sequences, highlighting their efficiency under constrained iteration budgets. Compared to OmeGA, our LLM-based methods often identify high-quality solutions earlier and continue refining them through iterative improvement. This observation supports the hypothesis that LLMs can perform efficient in-context learning, rapidly adapting and optimizing based on previously explored sequences. The only exception occurs in the Automobile Brake System DSM, where OmeGA achieves the optimal solution from the very beginning due to the relatively simple structure of the DSM’s strongly connected components. In this case, the block decomposition provided by the strongly connected components drastically reduces the search space, enabling the genetic algorithm with local search to immediately converge to the global optimum.

Moreover, the experimental results further highlight the value of contextual domain knowledge in improving the convergence behavior. In Figures 4A (UCAV), 4C (Heat Exchanger), and 4D (Brake System), the version of our method that incorporates domain knowledge (red line) consistently outperforms the variant without such knowledge (gray line), both in terms of convergence speed and final solution quality. This indicates that LLMs are capable of effectively combining semantic reasoning with structural optimization to solve CO problems more intelligently. A noteworthy exception appears in Figure 4B (Microfilm Cartridge), where both variants of the LLM-based method reach the optimal solution in the first step. As a result, the two lines overlap entirely, reflecting identical performance in this case. The findings indicate that for some DSMs with relatively less complexity, a small number of attempts may be sufficient for the LLM to infer optimal solutions.

\begin{figure}[H]
	\centering
	\includegraphics[width=15cm]{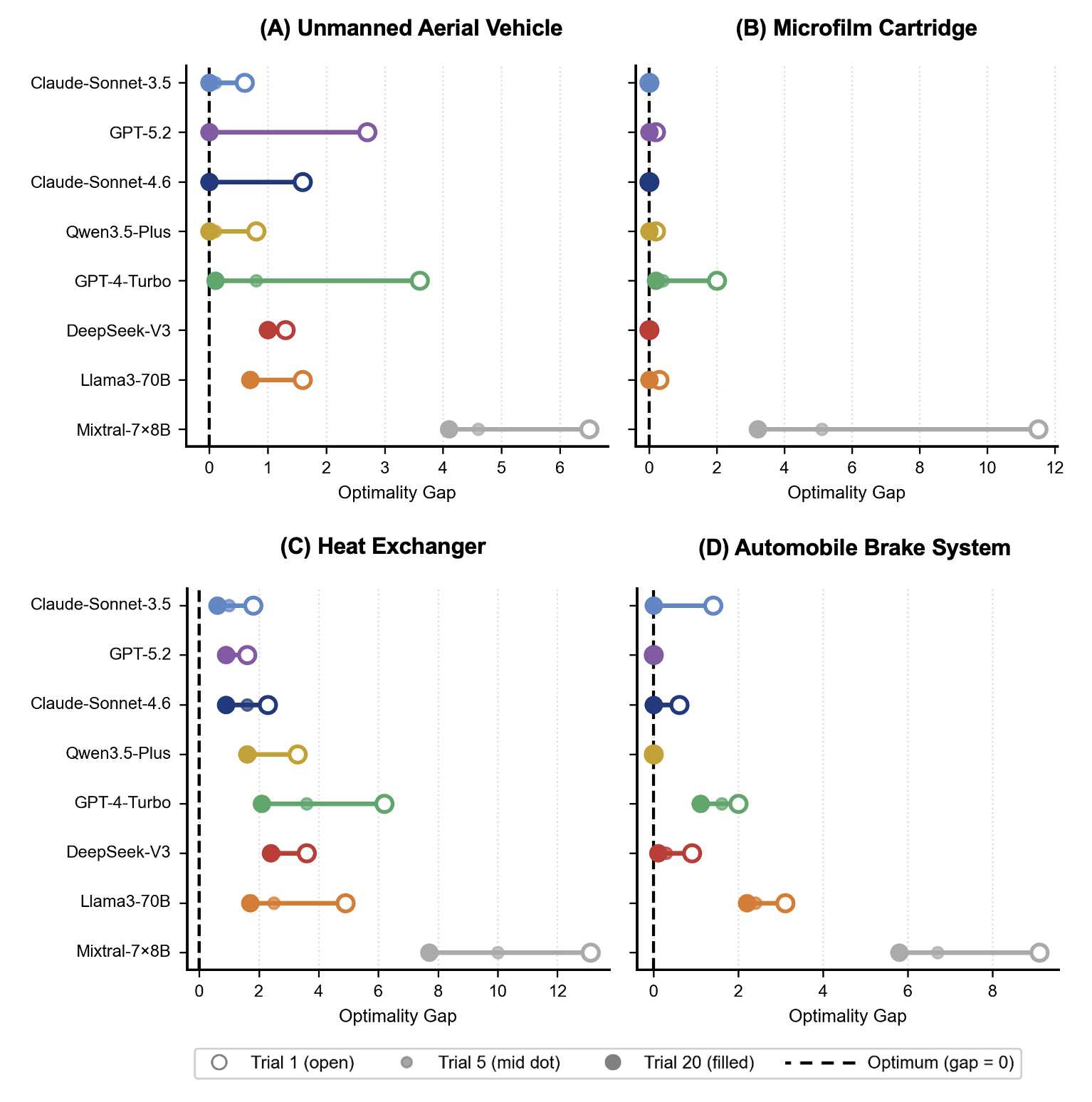}
	\caption{Optimality Gap of Different Backbone LLMs}
	\label{fig:fig5}
\end{figure}

\subsection{Impact of Backbone LLM Selection}

To evaluate the generalizability of our framework across different backbone LLMs, we tested eight models: \textit{Claude-Sonnet-3.5}, \textit{Claude-Sonnet-4.6}, \textit{GPT-5.2}, \textit{GPT-4-Turbo}, \textit{DeepSeek-V3}, \textit{Qwen-3.5-Plus}, \textit{Llama3-70B}, and \textit{Mixtral-7x8B}. Each model was evaluated under single-trial, 5-trial, and 20-trial settings with contextual domain knowledge included. Results are summarized in terms of optimality gap across all four DSM cases in Figure 5, with detailed numerical results provided in Appendix 3.

As shown in Figure 5, the top-tier proprietary models, including \textit{Claude-Sonnet-3.5}, \textit{Claude-Sonnet-4.6}, and \textit{GPT-5.2}, consistently achieve zero or near-zero optimality gaps across all four DSM cases, particularly under the 5-trial and 20-trial settings. \textit{Claude-Sonnet-3.5}, which serves as the primary backbone in our main experiments, achieves optimal results on three of four cases under 20 trials with no variance, and delivers competitive performance even under the single-trial setting, where it already outperforms most stochastic baselines that require thousands of evaluations. Mid-tier models, including \textit{DeepSeek-V3}, \textit{Qwen-3.5-Plus}, \textit{GPT-4-Turbo}, and \textit{Llama3-70B}, exhibit moderate performance with optimality gaps generally narrowing as the number of trials increases, making them viable alternatives when computational budget or deployment constraints are a concern. \textit{Mixtral-7x8B} consistently shows the largest optimality gaps across all cases and trial settings. No single open-source model dominates uniformly across all cases. \textit{DeepSeek-V3}, for instance, performs competitively on the Microfilm Cartridge and Brake System cases, while \textit{Qwen-3.5-Plus} shows stronger results on the UCAV case. This variability suggests that when open-source LLMs are adopted, a routing mechanism that dynamically selects models based on problem characteristics may further improve overall performance \cite{Hu2024}.

The results also confirm the critical role of contextual domain knowledge across all backbone LLMs. Incorporating domain knowledge consistently reduces the optimality gap compared to the knowledge-ablated setting, as detailed in Appendix 3. For \textit{Claude-Sonnet-3.5}, the gap between with-knowledge and without-knowledge settings is similarly consistent: on the Heat Exchanger case under 20 trials, the score improves from 4.1±0.3 to 3.6±0.5 when domain knowledge is incorporated. In addition, \textit{DeepSeek-V3} achieves 3.1±0.3 on the Brake System DSM under 20 trials with knowledge, compared to 4.0±1.3 without, a pattern observed consistently across models. This supports our earlier analysis, highlighting that incorporating contextual semantic reasoning significantly enhances solution quality and convergence performance in DSM optimization. This effect is especially notable in weaker models, where the inclusion of domain knowledge appears to offset some of the limitations in their underlying reasoning ability.

\subsection{Scalability to Larger Problem Instances and Alternative Objectives}

To assess the scalability of the proposed framework, we evaluated the three latest frontier LLMs on a larger case: an AW101 Helicopter DSM comprising 19 nodes and 109 edges \cite{Clarkson2004,Eckert2004}, substantially denser than the four cases examined in prior sections (12--17 nodes, 32--47 edges). As problem size grows, the input context fed to the LLM also expands, as longer candidate sequences must be encoded as few-shot examples in each prompt. Following the same evaluation protocol as in Section 5.1, results are reported in terms of optimality gap over 10 independent runs with contextual domain knowledge included. As shown in Figure 6, all three models show a consistent reduction in optimality gap as the number of trials increases, demonstrating that the iterative refinement mechanism remains effective at larger scale. \textit{Claude-Sonnet-4.6} achieves the lowest optimality gap of 1.7 under the 20-trial setting, indicating strong adaptability to the increased problem complexity. \textit{GPT-5.2} follows with a gap of 4.3, while \textit{Qwen-3.5-Plus} exhibits slower convergence, reaching a gap of 13.8 under the same setting, suggesting that the capacity to reason over longer input contexts varies across frontier models. It is worth noting that larger and more complex problem instances inherently benefit from additional iterations to further close the optimality gap; the 20-trial budget adopted here is maintained solely for consistency with the main experiments. At the same time, increased trial counts and longer input contexts directly translate to higher computational cost. Future research may explore adaptive strategies for balancing iteration budget, context length, and solution quality in LLM-based optimization.

To further examine whether the proposed framework generalizes to alternative optimization objectives, we evaluated three frontier LLMs using a weighted feedback metric \cite{Eppinger1994,Gebala1991}. These three models, \textit{Claude-Sonnet-4.6}, \textit{GPT-5.2}, and \textit{Qwen-3.5-Plus}, were selected as representative performers across different tiers based on the results of Section 5.3, as some earlier models are no longer accessible via API at the time of this evaluation. Formally, the score is computed as:

$$
\text{Score}(s) = \sum_{i=1}^{n} \sum_{j=i+1}^{n} a_{ij} \cdot (j - i)
$$

where a larger span between dependent nodes incurs a higher penalty, capturing the severity of rework rather than merely its frequency. Adopting this objective requires only a targeted modification to the prompt: the scoring function definition is updated to reflect the weighted criterion, while all other components of the framework, including the topological representation, contextual knowledge, and sampling strategy, remain unchanged. Baseline GA results are obtained under the same parameter settings as in the main experiments, as described in Appendix 2.

As shown in Table 3, the LLM-based methods retain strong optimization capability under this alternative objective. \textit{Claude-Sonnet-4.6} achieves the optimum on the Automobile Brake System (4.0±0.0) and closely approaches it on the UCAV case, performing comparably to GA-Exploration. \textit{GPT-5.2} delivers competitive results on most cases but shows a larger gap on the Heat Exchanger DSM, reflecting the added difficulty introduced by span-sensitive penalization. \textit{Qwen-3.5-Plus} reaches the optimum on the Microfilm Cartridge case (13.0±0.0) but lags on the remaining cases, consistent with its convergence behavior observed in prior sections. Overall, these results confirm that the proposed framework can be readily adapted to alternative combinatorial objectives with minimal modification, demonstrating its flexibility beyond the primary experimental setting. How to better convey complex objectives through prompt design to enhance LLM comprehension remains a promising direction for future work.

\begin{figure}[H]
	\centering
	\includegraphics[width=14cm]{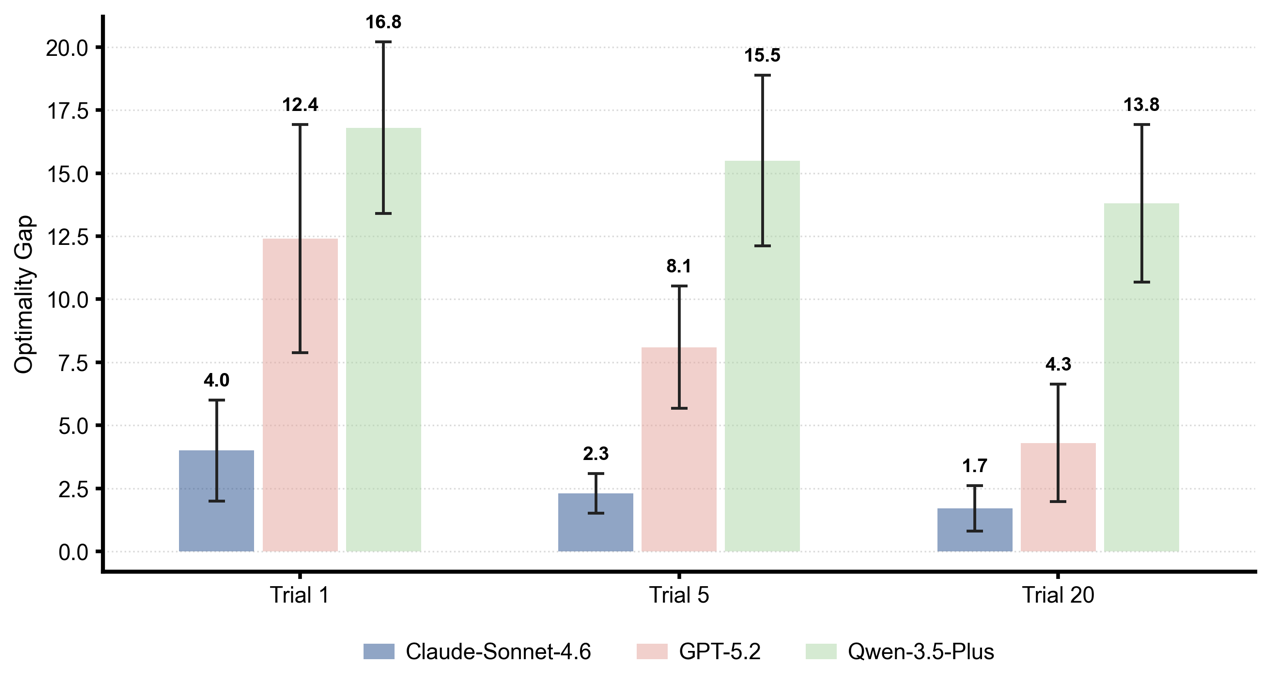}
	\caption{Scalability Evaluation on a Larger and Denser DSM Case (AW101 Helicopter DSM, Measure: Feedback Loops)}
	\label{fig:fig6}
\end{figure}

\begin{table}[H]
    \caption{Performance Comparison Under the Weighted Feedback Objective}
    \centering
    \renewcommand{\arraystretch}{1.3}
    \setlength{\tabcolsep}{6pt}
    \begin{tabular}{l m{2.3cm}<{\centering} m{2.3cm}<{\centering} m{2.3cm}<{\centering} m{2.3cm}<{\centering}}
        \toprule
        \textbf{Methods} & \multicolumn{2}{c}{\textbf{Activity-Based DSMs}} & \multicolumn{2}{c}{\textbf{Parameter-Based DSMs}} \\
        \cmidrule(lr){2-3} \cmidrule(lr){4-5}
        & \textbf{Unmanned Aerial Vehicle} & \textbf{Microfilm Cartridge} & \textbf{Heat Exchanger} & \textbf{Automobile Brake System} \\
        \midrule
        \textbf{LLM-based Methods (20 Trials)} & & & & \\
        \textit{Claude-Sonnet-4.6} & 24.6±0.8 & 13.2±0.6 & 11.0±2.1 & \textbf{4.0±0.0} \\
        \textit{GPT-5.2}           & 25.8±0.7 & 13.6±0.9 & 10.4±1.1 & 4.1±0.3 \\
        \textit{Qwen-3.5-Plus}     & 25.1±0.3 & \textbf{13.0±0.0} & 14.1±2.3 & 4.1±0.3 \\
        \midrule
        \textbf{Benchmark Methods} & & & & \\
        GA-Exploration@500  & 29.8±3.4 & 19.7±4.9 & 24.7±7.5 & 10.3±3.1 \\
        GA-Exploration@10K  & \textbf{24.0±0.0} & 13.8±1.6 & \textbf{8.0±1.2}  & 4.2±0.6 \\
        GA-Balance@500      & 33.2±6.8 & 28.7±6.4 & 27.9±4.7 & 16.4±4.8 \\
        GA-Balance@10K      & 24.2±0.6 & 13.4±1.2 & 8.8±1.2  & 5.4±2.0 \\
        GA-Exploitation@500 & 52.1±8.6 & 51.7±11.4 & 56.6±14.1 & 41.2±7.8 \\
        GA-Exploitation@10K & 26.5±2.5 & 17.8±3.1 & 17.5±6.1 & 8.3±2.8 \\
        \midrule
        \textbf{Optimum}    & 24       & 13        & 7         & 4 \\
        \bottomrule
    \end{tabular}
    \label{tab:table3}
\end{table}

\subsection{Practical Implications for Engineering Design}

So far, the experimental results have demonstrated that our LLM-based combinatorial optimization method is effective and efficient across various DSM cases. To illustrate the reasoning process of our method, we visualize a representative optimization trajectory randomly sampled from a single run of \textit{Claude-Sonnet-3.5} on the Heat Exchanger DSM (Figure 7), where the LLM progressively improves the node sequence over four iterations. In the first matrix randomly generated (Iteration 0), the initial node ordering yields 25 feedback loops, indicating a high level of dependency and inefficiency. By the first iteration, the LLM suggests a new configuration that already reduces the feedback loop count to 5. The third and final iteration further optimizes the sequence, achieving the minimal known value of 4 feedback loops. Throughout the optimization process, upstream nodes such as “Requirements” (N1), “Prognosis” (N3), and “Type of Concept” (N9) are consistently placed early, aligning with their role in initiating design decisions. Downstream elements like “Tool Manufacturing” (N17) and “Prototype Series” (N12) appear later in the sequence, preserving a logical and efficient flow of information. In practice, such visualization of the optimization trajectory enables us to analyze and trace the reasoning behind the LLM’s decisions, thereby enhancing the interpretability of the proposed method.

\begin{figure}[H]
	\centering
	\includegraphics[width=16.5cm]{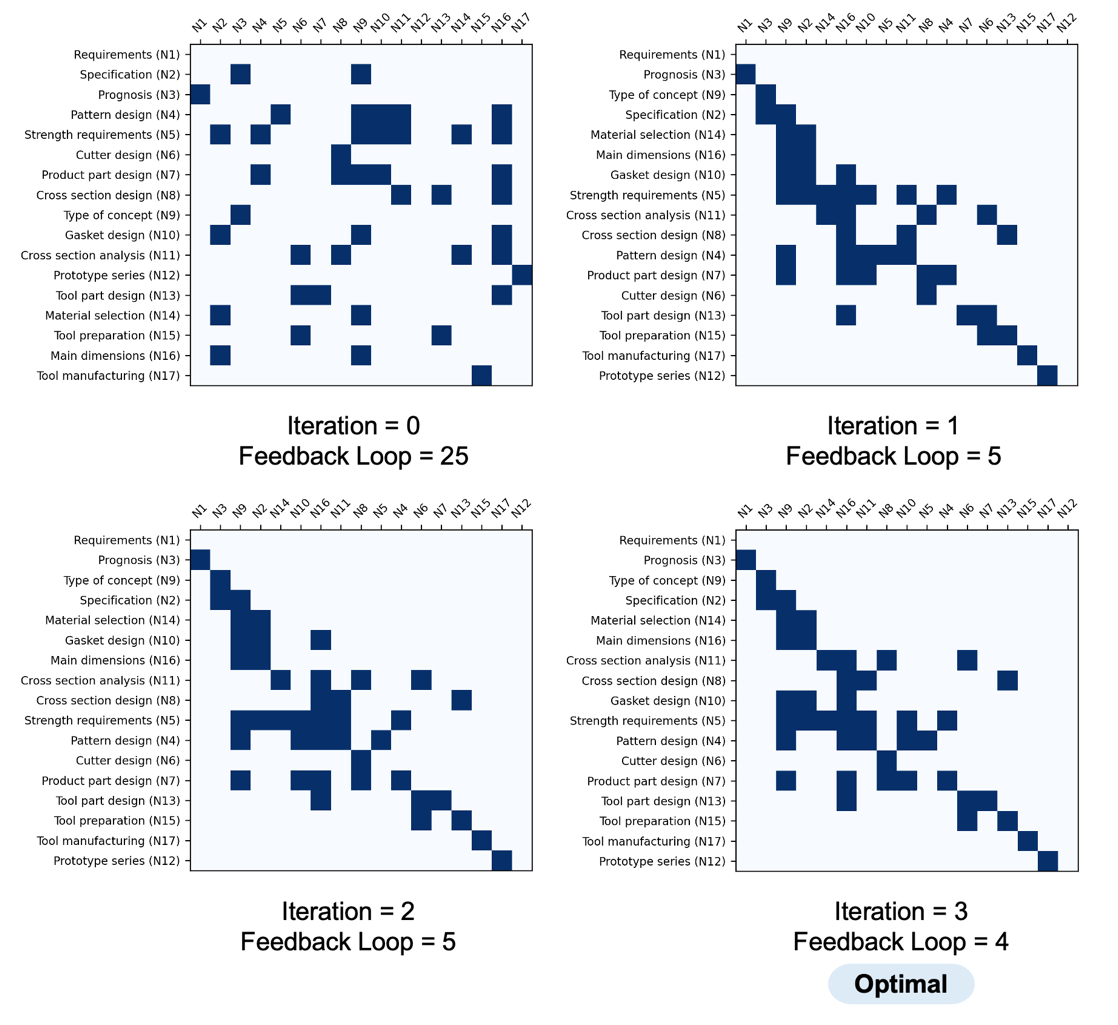}
	\caption{A Representative Optimization Trajectory for the Heat Exchanger DSM}
	\label{fig:fig7}
\end{figure}

Notably, for any given DSM, multiple optimal solutions may exist \cite{Eppinger2012}. With the proposed method, users can either select from the top candidates in the final solution base or rerun the method with different seeds to generate diverse and high-quality alternatives. This flexibility is particularly valuable in engineering contexts, where different project goals or constraints may influence the selection of one optimal solution over others.

Taken together, the proposed method can directly contribute to design planning and system-level coordination in engineering design practice. For example, in activity-based DSMs, it supports early-phase decision-making by suggesting task sequences that mitigate rework and streamline communication. In parameter-based DSMs, it reduces technical complexity by identifying sequencing patterns that minimize feedback loops. By leveraging both system structure information and domain knowledge, our method enables the LLM to reason in alignment with engineering contexts, rather than relying solely on statistical correlations.

While this study focuses on sequencing problems within DSMs, the proposed approach can be generalized to a broader class of CO problems \cite{Burggrf2024,Koh2015,Luo2015,Sarica2019,Sinha2018} in engineering and beyond. A more generalized framework is illustrated in Figure 8. During the engineering design process, systems or tasks can often be formulated as iterative optimization problems over formal representations, where multimodal information (such as task descriptions, parameter interactions, or architectural constraints encoded in text, diagrams, or code format) is encoded and integrated. The framework captures this pattern by transforming such information into a unified structured input and integrating this input with objective evaluation in a closed-loop reasoning cycle, enabling generation-based optimization using AI foundation models.

\begin{figure}[H]
	\centering
	\includegraphics[width=16.5cm]{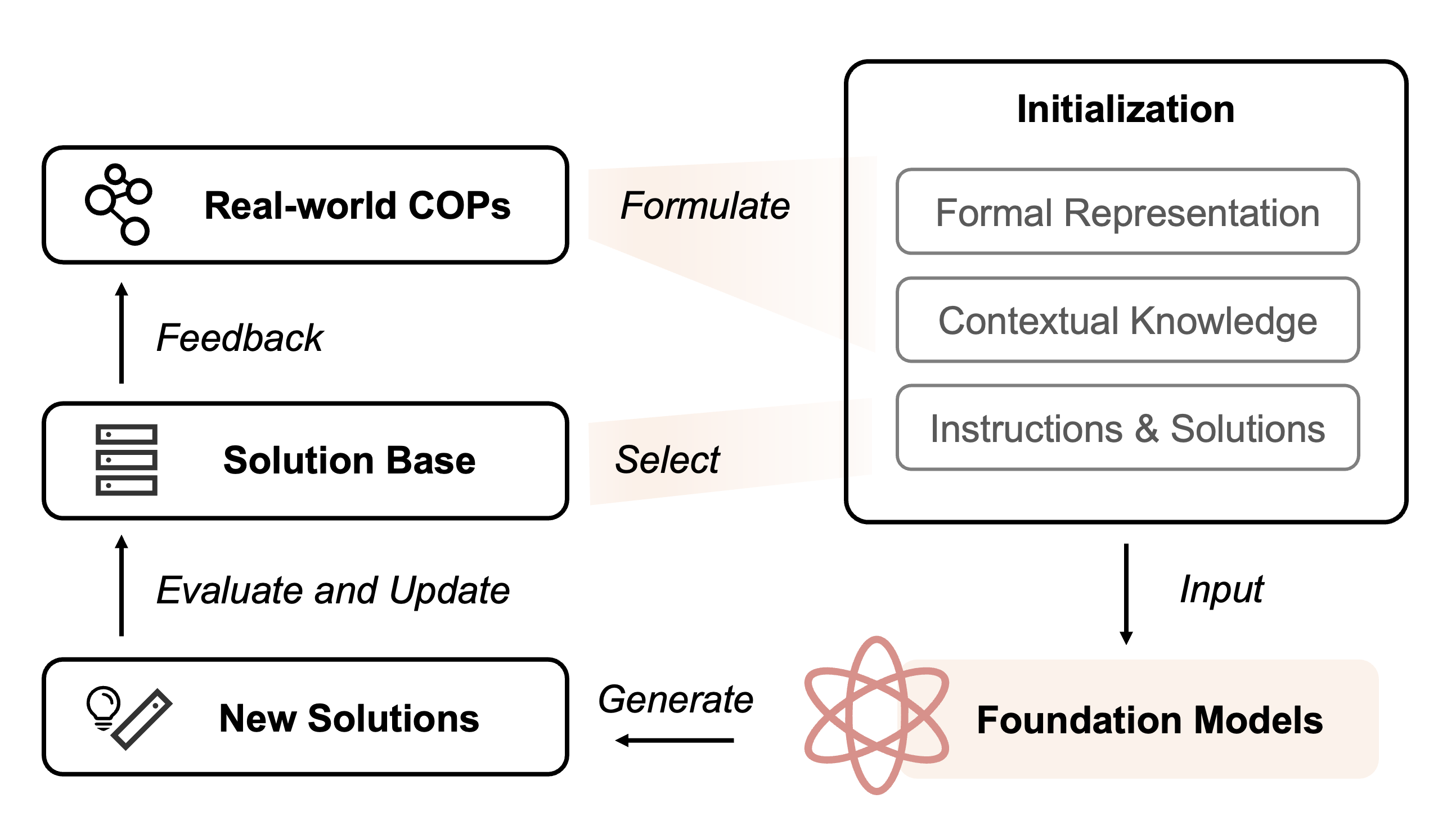}
	\caption{A Generalized Foundation Model-based Framework for Combinatorial Optimization}
	\label{fig:fig8}
\end{figure}

\section{Limitations and Future Work}

While the proposed method demonstrates strong performance, several limitations remain. First, DSMs in practice are highly diverse, not only in terms of size and structural complexity, but also in their representation. Beyond binary DSMs, weighted DSMs, multi-domain DSMs, and other enriched forms are common, where edge strength, type, or cross-domain dependencies play a critical role \cite{Eppinger2012}. Future work should evaluate the LLM-based methodology on a broader spectrum of DSM types to better reflect real engineering settings. Second, the scalability of the method to large-scale DSMs remains an open question. As system size grows, both computational performance and solution quality may be affected. While a preliminary evaluation on a larger case (the AW101 Helicopter DSM) demonstrates that the framework retains meaningful optimization capability at increased scale, the behavior of the method on even larger and more structurally complex networks remains uncertain. Further work is needed to systematically assess its efficiency and convergence under such conditions. Third, while this study primarily adopts the minimization of feedback loops as the sequencing objective, we also present a preliminary evaluation using a weighted feedback metric in Section 5.4. Future work should extend testing to a broader range of objectives, including concurrency, project cost and duration, and modularity \cite{Baldwin2000,Gebala1991,kusiak1993decomposition,Steward1981}, as well as different classes of combinatorial optimization problems beyond sequencing, such as modularization and change propagation analysis \cite{Burggrf2024,Koh2015,Luo2015,Sarica2019,Sinha2018}. In addition, how to better convey complex weighted objectives through prompt design to enhance LLM comprehension remains a promising direction.

Beyond performance and scalability, improving the interpretability of the optimization process represents another important direction for future research. In this paper, we presented a preliminary case illustrating how the LLM refines the DSM through a sequence of iterative updates, offering an initial window into the model's reasoning process. However, the underlying decision-making and reasoning mechanisms of LLMs still remain relatively opaque. Researchers may explore more effective techniques for LLM interpretability, such as integrating complementary models, enabling self-explanation, and visualizing decision pathways \cite{Singh2024}. Additionally, the current framework primarily relies on textual representations of domain knowledge and system structures. While effective, this may limit its ability to fully leverage non-textual modalities such as diagrams, tables, or images commonly used in engineering practice \cite{jiang2021deriving,jiang2022data}. In future work, we will explore Multimodal Large Language Models (MLLMs) to incorporate visual or tabular information \cite{Li2024}, enriching the reasoning input and expanding the framework’s applicability to more complex engineering workflows.

In addition, DSM methods have already been incorporated into commercial tools such as Loomeo (https://loomeo.com/) and Lattix (https://www.lattix.com/), which provide standardized DSM representations, built-in objectives, and user-friendly interfaces. Future studies may explore the comparison between our proposed method and these commercial tools. Moreover, building on this foundation, LLM-based optimization could be integrated into such platforms to strengthen their adaptability and intelligence. A potential implementation would involve using the existing DSM representation layer and evaluation functions as the infrastructure, while adding an LLM-based optimization module that generates and iteratively refines candidate sequences through prompt templates (refer to Appendix 1). The outputs from the LLM could then be automatically validated against the software’s scoring functions, and retrieval-augmented generation could be employed to incorporate external knowledge sources (e.g., design guidelines, prior project documentation) to enrich the reasoning process.

\section{Conclusion}

In this paper, we proposed a knowledge-informed optimization approach based on LLMs for solving DSM sequencing, a representative combinatorial optimization problem in engineering design. By integrating structural network information with contextual domain knowledge, the approach exploits the reasoning capabilities of LLMs to iteratively generate and refine candidate solutions. Experiments on various DSM cases show that the proposed method outperforms both stochastic and deterministic baselines in terms of convergence speed and solution quality. Further evaluations demonstrate that the framework scales to larger and more complex problem instances, and can be readily adapted to alternative optimization objectives with minimal modification. The consistent improvements observed when domain knowledge is incorporated highlight the benefit of combining semantic insight with mathematical structure. Overall, this study demonstrates that LLMs provide a promising foundation for advancing engineering design optimization and decision support, particularly in complex and knowledge-intensive engineering contexts.



\bibliographystyle{unsrtnat}
\bibliography{references}

@misc{meta2024llama,
   author = {Meta},
   title = {Introducing Meta Llama 3: The most capable openly available LLM to date},
   year = {2024},
   note = {https://ai.meta.com/blog/meta-llama-3/},
}

@misc{openai2024gpt4turbo,
   author = {OpenAI},
   title = {GPT-4 Turbo},
   year = {2024},
   note = {https://platform.openai.com/docs/models/gpt-4-and-gpt-4-turbo},
}

@misc{anthropic2024claude,
   author = {Anthropic},
   title = {Introducing computer use, a new Claude 3.5 Sonnet, and Claude 3.5 Haiku},
   year = {2024},
   note = {Accessed Oct 2024, https://www.anthropic.com/news/3-5-models-and-computer-use},
}

@article{Jiang2025,
   author = {Shuo Jiang and Weifeng Li and Yuping Qian and Yangjun Zhang and Jianxi Luo},
   journal = {Advanced Engineering Informatics},
   pages = {103312},
   publisher = {Elsevier},
   title = {AutoTRIZ: Automating engineering innovation with TRIZ and large language models},
   volume = {65},
   year = {2025}
}

@article{Steward1981,
   author = {Donald V Steward},
   issue = {3},
   journal = {IEEE transactions on Engineering Management},
   pages = {71-74},
   publisher = {IEEE},
   title = {The design structure system: A method for managing the design of complex systems},
   year = {1981}
}

@article{Black1990,
   author = {Thomas Andrew Black and Charles H Fine and Emanuel M Sachs and others},
   publisher = {Sloan School of Management, Massachusetts Institute of Technology},
   title = {A method for systems design using precedence relationships: An application to automotive brake systems},
   year = {1990}
}

@phdthesis{Sequeira1991,
   author = {Michele Wanda Sequeira},
   school = {Massachusetts Institute of Technology},
   title = {Use of the design structure matrix in the improvement of an automobile development process},
   year = {1991}
}

@article{Dietzenbacher1992,
   author = {Erik Dietzenbacher},
   issue = {4},
   journal = {Economic Modelling},
   pages = {419-437},
   publisher = {Elsevier},
   title = {The measurement of interindustry linkages: Key sectors in the Netherlands},
   volume = {9},
   year = {1992}
}

@phdthesis{Osborne1993,
   author = {Sean M Osborne},
   school = {Massachusetts Institute of Technology},
   title = {Product development cycle time characterization through modeling of process iteration},
   year = {1993}
}

@article{Kusiak1993,
   author = {Andrew Kusiak and Juite Wang},
   issue = {4},
   journal = {The International Journal Of Production Research},
   pages = {753-769},
   publisher = {Taylor \& Francis},
   title = {Efficient organizing of design activities},
   volume = {31},
   year = {1993}
}

@article{Eppinger1994,
   author = {Steven D Eppinger and Daniel E Whitney and Robert P Smith and David A Gebala},
   journal = {Research in engineering design},
   pages = {1-13},
   publisher = {Springer},
   title = {A model-based method for organizing tasks in product development},
   volume = {6},
   year = {1994}
}

@book{Rogers1996,
   author = {James L Rogers and Collin M McCulley and Christina L Bloebaum},
   publisher = {Springer},
   title = {Integrating a genetic algorithm into a knowledge-based system for ordering complex design processes},
   year = {1996}
}

@article{Smith1997,
   author = {Robert P Smith and Steven D Eppinger},
   issue = {8},
   journal = {Management Science},
   pages = {1104-1120},
   publisher = {INFORMS},
   title = {A predictive model of sequential iteration in engineering design},
   volume = {43},
   year = {1997}
}

@phdthesis{Browning1998,
   author = {Tyson R Browning},
   school = {Massachusetts Institute of Technology, Sloan School of Management~…},
   title = {Modeling and analyzing cost, schedule, and performance in complex system product development},
   year = {1998}
}

@article{Hameri1999,
   author = {A-P Hameri},
   issue = {6},
   journal = {International Journal of Production Research},
   pages = {1319-1336},
   publisher = {Taylor \& Francis},
   title = {Document viewpoint on one-of-a-kind delivery process},
   volume = {37},
   year = {1999}
}

@inproceedings{Amen1999,
   author = {Rafael Amen and Ingvar Rask and Staffan Sunnersjö},
   booktitle = {International Design Engineering Technical Conferences and Computers and Information in Engineering Conference},
   pages = {1165-1174},
   title = {Matching design tasks to knowledge-based software tools: When intuition does not suffice},
   volume = {19715},
   year = {1999}
}

@article{Clarkson2000,
   author = {Peter John Clarkson and James Robert Hamilton},
   journal = {Research in Engineering Design},
   pages = {18-38},
   publisher = {Springer},
   title = {‘Signposting’, a parameter-driven task-based model of the design process},
   volume = {12},
   year = {2000}
}

@article{Ahmadi2001,
   author = {Reza Ahmadi and Thomas A Roemer and Robert H Wang},
   issue = {3},
   journal = {European Journal of Operational Research},
   pages = {539-558},
   publisher = {Elsevier},
   title = {Structuring product development processes},
   volume = {130},
   year = {2001}
}

@article{Browning2001,
   author = {Tyson R Browning},
   issue = {3},
   journal = {IEEE Transactions on Engineering management},
   pages = {292-306},
   publisher = {IEEE},
   title = {Applying the design structure matrix to system decomposition and integration problems: a review and new directions},
   volume = {48},
   year = {2001}
}

@article{Estrada2005,
   author = {Ernesto Estrada and Juan A Rodriguez-Velazquez},
   issue = {5},
   journal = {Physical Review E—Statistical, Nonlinear, and Soft Matter Physics},
   pages = {56103},
   publisher = {APS},
   title = {Subgraph centrality in complex networks},
   volume = {71},
   year = {2005}
}

@article{Karniel2009,
   author = {Arie Karniel and Yoram Reich},
   issue = {4},
   journal = {IEEE Transactions on Engineering Management},
   pages = {636-649},
   publisher = {IEEE},
   title = {From DSM-based planning to design process simulation: a review of process scheme logic verification issues},
   volume = {56},
   year = {2009}
}

@article{Estrada2010,
   author = {Ernesto Estrada and Desmond J Higham},
   issue = {4},
   journal = {SIAM review},
   pages = {696-714},
   publisher = {SIAM},
   title = {Network properties revealed through matrix functions},
   volume = {52},
   year = {2010}
}

@inproceedings{Choi2011,
   author = {Young Mi Choi and others},
   booktitle = {DS 68-3: Proceedings of the 18th International Conference on Engineering Design (ICED 11), Impacting Society through Engineering Design, Vol. 3: Design Organisation and Management, Lyngby/Copenhagen, Denmark, 15.-19.08. 2011},
   pages = {116-122},
   title = {Effective scheduling of user input during the design process},
   year = {2011}
}

@article{Crofts2011,
   author = {Jonathan J Crofts and Desmond J Higham},
   issue = {4},
   journal = {Internet Mathematics},
   pages = {233-254},
   publisher = {Taylor \& Francis},
   title = {Googling the brain: Discovering hierarchical and asymmetric network structures, with applications in neuroscience},
   volume = {7},
   year = {2011}
}

@book{Korte2011,
   author = {Bernhard H Korte and Jens Vygen and B Korte and J Vygen},
   publisher = {Springer},
   title = {Combinatorial optimization},
   volume = {1},
   year = {2011}
}

@article{Qian2011,
   author = {Yanjun Qian and Jun Lin and Thong Ngee Goh and Min Xie},
   issue = {4},
   journal = {IEEE Transactions on Engineering Management},
   pages = {688-705},
   publisher = {IEEE},
   title = {A novel approach to DSM-based activity sequencing problem},
   volume = {58},
   year = {2011}
}

@article{Lin2012,
   author = {Jun Lin and Yanjun Qian and Ali A Yassine and Wentian Cui},
   issue = {23},
   journal = {International journal of production research},
   pages = {7012-7025},
   publisher = {Taylor \& Francis},
   title = {A fuzzy approach for sequencing interrelated activities in a DSM},
   volume = {50},
   year = {2012}
}

@article{Pasqual2012,
   author = {Michael C Pasqual and Olivier L de Weck},
   journal = {Research in Engineering Design},
   pages = {305-328},
   publisher = {Springer},
   title = {Multilayer network model for analysis and management of change propagation},
   volume = {23},
   year = {2012}
}

@book{Eppinger2012,
   author = {Steven D Eppinger and Tyson R Browning},
   publisher = {MIT press},
   title = {Design structure matrix methods and applications},
   year = {2012}
}

@article{Fortin2012,
   author = {Félix-Antoine Fortin and François-Michel De Rainville and Marc-André Gardner Gardner and Marc Parizeau and Christian Gagné},
   issue = {1},
   journal = {The Journal of Machine Learning Research},
   pages = {2171-2175},
   publisher = {JMLR. org},
   title = {DEAP: Evolutionary algorithms made easy},
   volume = {13},
   year = {2012}
}

@article{MacCormack2012,
   author = {Alan MacCormack and Carliss Baldwin and John Rusnak},
   issue = {8},
   journal = {Research policy},
   pages = {1309-1324},
   publisher = {Elsevier},
   title = {Exploring the duality between product and organizational architectures: A test of the “mirroring” hypothesis},
   volume = {41},
   year = {2012}
}

@book{Pinedo2012,
   author = {Michael L Pinedo},
   publisher = {Springer},
   title = {Scheduling},
   volume = {29},
   year = {2012}
}

@article{Koh2015,
   author = {Edwin C Y Koh and Armin Förg and Matthias Kreimeyer and Markus Lienkamp},
   journal = {Research in Engineering Design},
   pages = {337-353},
   publisher = {Springer},
   title = {Using engineering change forecast to prioritise component modularisation},
   volume = {26},
   year = {2015}
}

@book{Eppinger2016,
   author = {Steven D Eppinger and Karl Ulrich},
   publisher = {McGraw-Hill New York},
   title = {Product design and development},
   year = {2016}
}

@article{Sinha2018,
   author = {Kaushik Sinha and Eun Suk Suh},
   journal = {Research in Engineering Design},
   pages = {123-141},
   publisher = {Springer},
   title = {Pareto-optimization of complex system architecture for structural complexity and modularity},
   volume = {29},
   year = {2018}
}

@article{Attari-Shendi2019,
   author = {Milad Attari-Shendi and Mohammad Saidi-Mehrabad and Jafar Gheidar-Kheljani},
   issue = {6},
   journal = {IEEE Transactions on Engineering Management},
   pages = {2619-2633},
   publisher = {IEEE},
   title = {A comprehensive mathematical model for sequencing interrelated activities in complex product development projects},
   volume = {69},
   year = {2019}
}

@article{Ogura2019,
   author = {Masaki Ogura and Junichi Harada and Masako Kishida and Ali Yassine},
   journal = {Research in Engineering Design},
   pages = {435-452},
   publisher = {Springer},
   title = {Resource optimization of product development projects with time-varying dependency structure},
   volume = {30},
   year = {2019}
}

@inproceedings{Xidias2019,
   author = {Elias Xidias and Philip Azariadis},
   issue = {1},
   booktitle = {Proceedings of the Design Society: International Conference on Engineering Design},
   pages = {2853-2862},
   title = {Energy efficient motion design and task scheduling for an autonomous vehicle},
   volume = {1},
   year = {2019}
}

@article{Naseri2020,
   author = {Gita Naseri and Mattheos A G Koffas},
   issue = {1},
   journal = {Nature communications},
   pages = {2446},
   publisher = {Nature Publishing Group UK London},
   title = {Application of combinatorial optimization strategies in synthetic biology},
   volume = {11},
   year = {2020}
}

@article{Wei2022,
   author = {Jason Wei and Yi Tay and Rishi Bommasani and Colin Raffel and Barret Zoph and Sebastian Borgeaud and Dani Yogatama and Maarten Bosma and Denny Zhou and Donald Metzler and others},
   journal = {Transactions on Machine Learning Research},
   title = {Emergent abilities of large language models},
   year = {2022}
}

@article{Zhao2023,
   author = {Wayne Xin Zhao and Kun Zhou and Junyi Li and Tianyi Tang and Xiaolei Wang and Yupeng Hou and Yingqian Min and Beichen Zhang and Junjie Zhang and Zican Dong and others},
   issue = {2},
   journal = {arXiv preprint arXiv:2303.18223},
   title = {A survey of large language models},
   volume = {1},
   year = {2023}
}

@article{Zhu2023,
   author = {Qihao Zhu and Jianxi Luo},
   issue = {4},
   journal = {Journal of Computing and Information Science in Engineering},
   pages = {41003},
   publisher = {American Society of Mechanical Engineers},
   title = {Generative transformers for design concept generation},
   volume = {23},
   year = {2023}
}

@article{Hu2024,
   author = {Qitian Jason Hu and Jacob Bieker and Xiuyu Li and Nan Jiang and Benjamin Keigwin and Gaurav Ranganath and Kurt Keutzer and Shriyash Kaustubh Upadhyay},
   journal = {arXiv preprint arXiv:2403.12031},
   title = {Routerbench: A benchmark for multi-llm routing system},
   year = {2024}
}

@article{Picard2024,
   author = {Cyril Picard and Lyle Regenwetter and Amin Heyrani Nobari and Akash Srivastava and Faez Ahmed},
   journal = {arXiv preprint arXiv:2412.13281},
   title = {Generative Optimization: A Perspective on AI-Enhanced Problem Solving in Engineering},
   year = {2024}
}

@article{Burggrf2024,
   author = {Peter Burggräf and Johannes Wagner and Till Saßmannshausen and Tim Weißer and Ognjen Radisic-Aberger},
   issue = {2},
   journal = {Research in Engineering Design},
   pages = {215-237},
   publisher = {Springer},
   title = {AI-artifacts in engineering change management–a systematic literature review},
   volume = {35},
   year = {2024}
}

@article{Singh2024,
   author = {Chandan Singh and Jeevana Priya Inala and Michel Galley and Rich Caruana and Jianfeng Gao},
   journal = {arXiv preprint arXiv:2402.01761},
   title = {Rethinking interpretability in the era of large language models},
   year = {2024}
}

@inproceedings{Li2024,
   author = {Bohao Li and Yuying Ge and Yixiao Ge and Guangzhi Wang and Rui Wang and Ruimao Zhang and Ying Shan},
   booktitle = {Proceedings of the IEEE/CVF Conference on Computer Vision and Pattern Recognition},
   pages = {13299-13308},
   title = {Seed-bench: Benchmarking multimodal large language models},
   year = {2024}
}

@article{Liu2024deepseek,
   author = {Aixin Liu and Bei Feng and Bing Xue and Bingxuan Wang and Bochao Wu and Chengda Lu and Chenggang Zhao and Chengqi Deng and Chenyu Zhang and Chong Ruan and others},
   journal = {arXiv preprint arXiv:2412.19437},
   title = {Deepseek-v3 technical report},
   year = {2024}
}

@inproceedings{Liu2024eoh,
   author = {Fei Liu and Tong Xialiang and Mingxuan Yuan and Xi Lin and Fu Luo and Zhenkun Wang and Zhichao Lu and Qingfu Zhang},
   booktitle = {International Conference on Machine Learning},
   pages = {32201-32223},
   title = {Evolution of Heuristics: Towards Efficient Automatic Algorithm Design Using Large Language Model},
   year = {2024}
}

@article{Koh2024,
   author = {Edwin C Y Koh},
   journal = {Natural Language Processing Journal},
   pages = {100103},
   publisher = {Elsevier},
   title = {Auto-DSM: Using a Large Language Model to generate a Design Structure Matrix},
   volume = {9},
   year = {2024}
}

@article{Bordas2024,
   author = {Antoine Bordas and Pascal Le Masson and Maxime Thomas and Benoit Weil},
   issue = {4},
   journal = {Research in Engineering Design},
   pages = {427-443},
   publisher = {Springer},
   title = {What is generative in generative artificial intelligence? A design-based perspective},
   volume = {35},
   year = {2024}
}

@article{Wong2024,
   author = {Foo Shing Wong and David C Wynn},
   issue = {4},
   journal = {Research in Engineering Design},
   pages = {389-408},
   publisher = {Springer},
   title = {M-ARM: An automated systematic approach for generating new variant design options from an existing product family},
   volume = {35},
   year = {2024}
}

@article{Mei2024,
   author = {Aoran Mei and Guo-Niu Zhu and Huaxiang Zhang and Zhongxue Gan},
   journal = {IEEE Robotics and Automation Letters},
   publisher = {IEEE},
   title = {ReplanVLM: Replanning robotic tasks with visual language models},
   year = {2024}
}

@article{Alhijawi2024,
   author = {Bushra Alhijawi and Arafat Awajan},
   issue = {3},
   journal = {Evolutionary Intelligence},
   pages = {1245-1256},
   publisher = {Springer},
   title = {Genetic algorithms: Theory, genetic operators, solutions, and applications},
   volume = {17},
   year = {2024}
}

@article{Chang2024,
   author = {Yupeng Chang and Xu Wang and Jindong Wang and Yuan Wu and Linyi Yang and Kaijie Zhu and Hao Chen and Xiaoyuan Yi and Cunxiang Wang and Yidong Wang and others},
   issue = {3},
   journal = {ACM transactions on intelligent systems and technology},
   pages = {1-45},
   publisher = {ACM New York, NY},
   title = {A survey on evaluation of large language models},
   volume = {15},
   year = {2024}
}

@inproceedings{Liu2024llmaseo,
   author = {Shengcai Liu and Caishun Chen and Xinghua Qu and Ke Tang and Yew-Soon Ong},
   booktitle = {2024 IEEE Congress on Evolutionary Computation (CEC)},
   pages = {1-8},
   title = {Large language models as evolutionary optimizers},
   year = {2024}
}

@article{kusiak1993decomposition,
  title={Decomposition of the Design Process},
  author={Kusiak, A and Wang, J},
  journal={Journal of Mechanical Design},
  volume={115},
  pages={687},
  year={1993}
}

@article{Makatura2024,
   author = {Liane Makatura and Michael Foshey and Bohan Wang and Felix Hähnlein and Pingchuan Ma and Bolei Deng and Megan Tjandrasuwita and Andrew Spielberg and Crystal Elaine Owens and Peter Yichen Chen and Allan Zhao and Amy Zhu and Wil J Norton and Edward Gu and Joshua Jacob and Yifei Li and Adriana Schulz and Wojciech Matusik},
   journal = {An MIT Exploration of Generative AI},
   month = {3},
   note = {https://mit-genai.pubpub.org/pub/nmypmnhs},
   publisher = {MIT},
   title = {Large Language Models for Design and Manufacturing},
   year = {2024}
}

@article{Romera-Paredes2024,
   author = {Bernardino Romera-Paredes and Mohammadamin Barekatain and Alexander Novikov and Matej Balog and M Pawan Kumar and Emilien Dupont and Francisco J R Ruiz and Jordan S Ellenberg and Pengming Wang and Omar Fawzi and others},
   issue = {7995},
   journal = {Nature},
   pages = {468-475},
   publisher = {Nature Publishing Group UK London},
   title = {Mathematical discoveries from program search with large language models},
   volume = {625},
   year = {2024}
}

@inproceedings{Yang2024,
   author = {Chengrun Yang and Xuezhi Wang and Yifeng Lu and Hanxiao Liu and Quoc V Le and Denny Zhou and Xinyun Chen},
   booktitle = {The Twelfth International Conference on Learning Representations},
   title = {Large Language Models as Optimizers},
   year = {2024}
}

@article{Altus1996,
   author = {Stephen S Altus and Ilan M Kroo and Peter J Gage},
   doi = {10.1115/1.2826916},
   issn = {1050-0472},
   issue = {4},
   journal = {Journal of Mechanical Design},
   month = {9},
   pages = {486-489},
   title = {A Genetic Algorithm for Scheduling and Decomposition of Multidisciplinary Design Problems},
   volume = {118},
   url = {https://doi.org/10.1115/1.2826916},
   year = {1996}
}

@book{Baldwin2000,
   author = {Carliss Y Baldwin and Kim B Clark},
   publisher = {MIT press},
   title = {Design rules: The power of modularity},
   year = {2000}
}

@inproceedings{Gebala1991,
   author = {David A Gebala and Steven D Eppinger},
   booktitle = {International design engineering technical conferences and computers and information in engineering conference},
   pages = {227-233},
   title = {Methods for analyzing design procedures},
   volume = {7477},
   year = {1991}
}

@article{Meier2006,
   author = {Christoph Meier and Ali A Yassine and Tyson R Browning},
   doi = {10.1115/1.2717224},
   issn = {1050-0472},
   issue = {6},
   journal = {Journal of Mechanical Design},
   month = {9},
   pages = {566-585},
   title = {Design Process Sequencing With Competent Genetic Algorithms},
   volume = {129},
   url = {https://doi.org/10.1115/1.2717224},
   year = {2006}
}

@article{Luo2012,
   author = {Jianxi Luo and Carliss Y Baldwin and Daniel E Whitney and Christopher L Magee},
   issue = {6},
   journal = {Industrial and Corporate Change},
   pages = {1307-1335},
   publisher = {Oxford University Press},
   title = {The architecture of transaction networks: a comparative analysis of hierarchy in two sectors},
   volume = {21},
   year = {2012}
}

@article{Luo2015,
   author = {Jianxi Luo},
   issue = {4},
   journal = {Research in Engineering Design},
   pages = {355-371},
   publisher = {Springer},
   title = {A simulation-based method to evaluate the impact of product architecture on product evolvability},
   volume = {26},
   year = {2015}
}

@article{Sarica2019,
   author = {Serhad Sarica and Jianxi Luo},
   issue = {4},
   journal = {IEEE systems journal},
   pages = {3610-3618},
   publisher = {IEEE},
   title = {An infinite regress model of design change propagation in complex systems},
   volume = {13},
   year = {2019}
}

@article{Tarjan1972,
   author = {Robert Tarjan},
   issue = {2},
   journal = {SIAM journal on computing},
   pages = {146-160},
   publisher = {SIAM},
   title = {Depth-first search and linear graph algorithms},
   volume = {1},
   year = {1972}
}

@article{luo2019hierarchy,
  title={The hierarchy-niche model for supply networks},
  author={Luo, Jianxi and Whitney, Daniel E},
  journal={Production and Operations Management},
  volume={28},
  number={3},
  pages={668--681},
  year={2019},
  publisher={Sage Publications Sage CA: Los Angeles, CA}
}

@article{jiang2021deriving,
  title={Deriving design feature vectors for patent images using convolutional neural networks},
  author={Jiang, Shuo and Luo, Jianxi and Ruiz-Pava, Guillermo and Hu, Jie and Magee, Christopher L},
  journal={Journal of Mechanical Design},
  volume={143},
  number={6},
  pages={061405},
  year={2021},
  publisher={American Society of Mechanical Engineers}
}

@article{jiang2022data,
  title={Data-driven design-by-analogy: state-of-the-art and future directions},
  author={Jiang, Shuo and Hu, Jie and Wood, Kristin L and Luo, Jianxi},
  journal={Journal of Mechanical Design},
  volume={144},
  number={2},
  pages={020801},
  year={2022},
  publisher={American Society of Mechanical Engineers}
}

@article{Koh2026,
   author = {Edwin C Y Koh},
   journal = {Research in Engineering Design},
   pages = {13},
   publisher = {Springer},
   title = {From text to {DSM}: evaluating the impact of writing style and entity naming on {LLM}-based retrieval of asymmetrical indirect design dependencies},
   volume = {37},
   year = {2026}
}

@article{Clarkson2004,
   author = {Peter John Clarkson and Caroline Simons and Claudia Eckert},
   journal = {Journal of Mechanical Design},
   pages = {788-797},
   publisher = {ASME},
   title = {Predicting change propagation in complex design},
   volume = {126},
   year = {2004}
}

@article{Eckert2004,
   author = {Claudia Eckert and Peter John Clarkson and Walter Zanker},
   journal = {Research in Engineering Design},
   pages = {1-21},
   publisher = {Springer},
   title = {Change and customisation in complex engineering domains},
   volume = {15},
   year = {2004}
}

@article{Zhang2026,
   author = {Yang Zhang and Xinjing Wang and Zuochen Zhang and Jian Shen and Weifeng Li},
   journal = {Research in Engineering Design},
   pages = {14},
   publisher = {Springer},
   title = {Closed-loop integrated design framework for complex mechatronic system},
   volume = {37},
   year = {2026}
}

@article{JiangID2025,
   author = {Shuo Jiang and Min Xie and Fuyan Chen and Jiawei Ma and Jianxi Luo},
   journal = {Journal of Computing and Information Science in Engineering},
   title = {Intelligent Design 4.0: Paradigm Evolution Toward the Agentic {AI} Era},
   year = {2025}
}

\newpage
\section*{Appendix 1. Full prompts}

\subsection*{Prompt for the proposed approach:}

\begin{tcolorbox}[colback=gray!10, colframe=white, coltitle=black, boxrule=0.5mm, arc=0mm, left=2mm, right=2mm, top=1mm, bottom=1mm, width=\textwidth]
You are an expert in the domain of combinational optimization.
\\
\\
Please assist me to find an optimal sequential order that minimizes feedback cycles in the dependency network described below. Your task is to propose a new order that differs from previous attempts and has fewer feedback cycles than any listed.
\\
\\
\texttt{<Description of the Entire Network>} \texttt{\{network\_description\}} \texttt{</Description of the Entire Network>}\\
\texttt{<Nodes with Descriptions>} \texttt{\{node\_list\_with\_description\}} \texttt{</Nodes with Descriptions>}\\
\texttt{<Edges>} \texttt{\{edge\_list\}} \texttt{</Edges>}
\\
\\
Below are some previous sequential orders arranged in descending order of feedback cycles (lower is better): \texttt{\{selected\_historical\_solutions\}}
\\
\\
Please suggest a new order that:\\
- Is different from all prior orders.\\
- Has fewer feedback cycles than any previous order.\\
- Covers all nodes exactly once.\\
- Starts with \texttt{<order>} and ends with \texttt{</order>}.\\
- You can use the descriptions of nodes and networks to support your suggestion.
\\
\\
Output Format:\\
\texttt{<order> ...... </order>}
\\
\\
Please provide only the order and nothing else.
\end{tcolorbox}

\subsection*{Prompt for the proposed approach (without knowledge):}

\begin{tcolorbox}[colback=gray!10, colframe=white, coltitle=black, boxrule=0.5mm, arc=0mm, left=2mm, right=2mm, top=1mm, bottom=1mm, width=\textwidth]
You are an expert in the domain of combinational optimization.
\\
\\
Please assist me to find an optimal sequential order that minimizes feedback cycles in the dependency network described below. Your task is to propose a new order that differs from previous attempts and has fewer feedback cycles than any listed.
\\
\\
\texttt{<Nodes>} \texttt{\{node\_list\}} \texttt{</Nodes>}\\
\texttt{<Edges>} \texttt{\{edge\_list\}} \texttt{</Edges>}
\\
\\
Below are some previous sequential orders arranged in descending order of feedback cycles (lower is better): \texttt{\{selected\_historical\_solutions\}}
\\
\\
Please suggest a new order that:\\
- Is different from all prior orders.\\
- Has fewer feedback cycles than any previous order.\\
- Covers all nodes exactly once.\\
- Starts with \texttt{<order>} and ends with \texttt{</order>}.
\\
\\
Output Format:\\
\texttt{<order> ...... </order>}
\\
\\
Please provide only the order and nothing else.
\end{tcolorbox}

\subsection*{Example of \texttt{\{network\_description\}} in the UCAV design activity DSM case:}

\begin{tcolorbox}[colback=gray!10, colframe=white, coltitle=black, boxrule=0.5mm, arc=0mm, left=2mm, right=2mm, top=1mm, bottom=1mm, width=\textwidth]

\texttt{This network represents the dependency relationships among conceptual design activities for UCAV development at Boeing. Each node corresponds to a specific task or analysis, and directed edges indicate the prerequisite relationships between these tasks. Nodes: Each node is a task or analysis in the conceptual design process. Edges: Directed edges show the prerequisite relationships between these tasks and analyses.}

\end{tcolorbox}

\subsection*{Example of \texttt{\{node\_list\_with\_description\}} in the UCAV design activity DSM case:}

\begin{tcolorbox}[colback=gray!10, colframe=white, coltitle=black, boxrule=0.5mm, arc=0mm, left=2mm, right=2mm, top=1mm, bottom=1mm, width=\textwidth]

\texttt{[}\\
\texttt{\{'id': 'lzOtR', 'name': 'Create Configuration Concepts'\},}\\
\texttt{\{'id': 'yLlKi', 'name': 'Prepare UCAV Conceptual DR\&O'\},}\\
\texttt{\{'id': 'Swvi2', 'name': 'Prepare 3-View Drawing \& Geometry Data'\},}\\
\texttt{\{'id': 'CDcxF', 'name': 'Perform Weights Analyses \& Evaluation'\},}\\
\texttt{\{'id': '0KGDm', 'name': 'Perform Aerodynamics Analyses \& Evaluation'\},}\\
\texttt{\{'id': '4wHtv', 'name': 'Perform Multidisciplinary Analyses \& Evaluation'\},}\\
\texttt{\{'id': 'AgIBP', 'name': 'Prepare \& Distribute Choice Config. Data Set'\},}\\
\texttt{\{'id': 'gRtHi', 'name': 'Perform S\&C Characteristics Analyses \& Eval.'\},}\\
\texttt{\{'id': 'GV9RJ', 'name': 'Make Concept Assessment and Variant Decisions'\},}\\
\texttt{\{'id': 'I1j2m', 'name': 'Perform Performance Analyses \& Evaluation'\},}\\
\texttt{\{'id': 'Vzzm7', 'name': 'Perform Propulsion Analyses \& Evaluation'\},}\\
\texttt{\{'id': 'B0BFG', 'name': 'Perform Mechanical \& Electrical Analyses \& Eval.'\}}\\
\texttt{]}
\end{tcolorbox}

\subsection*{Example of \texttt{\{node\_list\}} in the UCAV design activity DSM case:}

\begin{tcolorbox}[colback=gray!10, colframe=white, coltitle=black, boxrule=0.5mm, arc=0mm, left=2mm, right=2mm, top=1mm, bottom=1mm, width=\textwidth]

\texttt{['lzOtR', 'yLlKi', 'Swvi2', 'CDcxF', '0KGDm', '4wHtv', 'AgIBP', 'gRtHi', 'GV9RJ', 'I1j2m', 'Vzzm7', 'B0BFG']}

\end{tcolorbox}

\subsection*{Example of \texttt{\{edge\_list\}} in the UCAV design activity DSM case:}

\begin{tcolorbox}[colback=gray!10, colframe=white, coltitle=black, boxrule=0.5mm, arc=0mm, left=2mm, right=2mm, top=1mm, bottom=1mm, width=\textwidth]

\texttt{[}\\
\texttt{\{'dependent': '0KGDm', 'predecessor': 'Swvi2'\},}\\
\texttt{\{'dependent': 'AgIBP', 'predecessor': 'lzOtR'\},}\\
\texttt{\{'dependent': '0KGDm', 'predecessor': 'yLlKi'\},}\\
\texttt{\{'dependent': 'Swvi2', 'predecessor': 'lzOtR'\},}\\
\texttt{\ldots}\\
\texttt{]}

\end{tcolorbox}

\subsection*{Example of \texttt{\{selected\_historical\_solutions\}} in the UCAV design activity DSM case:}

\begin{tcolorbox}[colback=gray!10, colframe=white, coltitle=black, boxrule=0.5mm, arc=0mm, left=2mm, right=2mm, top=1mm, bottom=1mm, width=\textwidth]

\texttt{[}\\
\texttt{\{'solution': 'lzOtR, yLlKi, GV9RJ, AgIBP, B0BFG, Vzzm7, Swvi2, CDcxF, 0KGDm, I1j2m, gRtHi, 4wHtv', 'score': 15.0\},}\\
\texttt{\{'solution': 'B0BFG, yLlKi, Vzzm7, lzOtR, Swvi2, CDcxF, AgIBP, 0KGDm, GV9RJ, I1j2m, gRtHi, 4wHtv', 'score': 13.0\},}\\
\texttt{\ldots}\\
\texttt{]}

\end{tcolorbox}

\raggedbottom

\section*{Appendix 2. Parameter Settings for Stochastic Baseline Methods}

In all three variants of the GA used for DSM sequencing tasks, we employed the following shared settings: the number of generations was set to 2,000, with a selection mechanism using tournament selection and mutation using shuffled indexes. Algorithms were implemented using the DEAP library with standard configurations for initial population generation. To ensure a fair comparison, all three GA variants and the OmeGA algorithm were initialized using the same seed solution as the proposed LLM-based methods. The different configurations for each GA setting are shown in Table 4 below. The detailed description of each parameter can be found in \cite{Fortin2012}.

\begin{table}[h!]
    \centering
    \caption{Parameter Settings for Three GA Variants}
    \renewcommand{\arraystretch}{1.3} 
    \setlength{\tabcolsep}{8pt} 
    \begin{tabular}{p{3.2cm} m{1.6cm}<{\centering} m{3cm}<{\centering} m{1.85cm}<{\centering} m{1.75cm}<{\centering} m{1.75cm}<{\centering}}
        \toprule
        & \textbf{Population} & \textbf{Individual Mutation Probability} & \textbf{Tournament Size} & \textbf{Crossover Probability} & \textbf{Mutation Probability} \\
        \midrule
        \textbf{Exploration-focused} & 50 & 0.05 & 5 & 0.6 & 0.4 \\
        \textbf{Exploitation-focused} & 10 & 0.01 & 20 & 0.9 & 0.1 \\
        \textbf{Balanced} & 20 & 0.02 & 10 & 0.7 & 0.3 \\
        \bottomrule
    \end{tabular}
    \label{tab:parameter_settings}
\end{table}

For the experimental setting of the Ordered Messy Genetic Algorithm (OmeGA) \cite{Meier2006}, the DSM was first decomposed into strongly connected components using the depth-first search algorithm \cite{Tarjan1972}, followed by topological sorting to determine inter-block ordering. Within each block, the GA was run with the same base settings as above but integrated with a 2-opt local search for further refinement. The population size per block was set to 50, the number of generations to 2,000, the tournament size to 4, crossover probability to 0.7, and mutation probability to 0.2. The local search was applied to the best individual in each generation with a maximum of 20 iterations. These parameter settings follow the literature with minor modifications \cite{Meier2006}.

\section*{Appendix 3. Detailed Results of the Backbone LLM Ablation}

The following table presents the detailed numerical results of the backbone LLM ablation study discussed in Section 5.3, where each entry reports the mean number of feedback loops over 10 independent runs, with $\pm$ indicating one standard deviation.
\newpage
\flushbottom

\thispagestyle{empty}
\vspace*{-1.7cm}
\begin{table}[H]
    \caption{Ablation on the Backbone LLM (Measure: Feedback Loops)}
    \centering
    \renewcommand{\arraystretch}{1.2}
    \setlength{\tabcolsep}{4pt}
    \begin{tabular}{p{1.55cm}<{\centering} m{2.7cm} m{1.7cm}<{\centering} p{2.15cm}<{\centering} p{2.15cm}<{\centering} p{2.15cm}<{\centering} p{2.15cm}<{\centering}}
        \toprule
        \multirow{2}{*}{\textbf{\# of Trial}} & \multirow{2}{*}{\textbf{Backbone LLM}} & \multirow{2}{*}{\textbf{Knowledge}}
        & \multicolumn{2}{c}{\textbf{Activity-Based DSMs}}
        & \multicolumn{2}{c}{\textbf{Parameter-Based DSMs}} \\
        \cmidrule(lr){4-5} \cmidrule(lr){6-7}
        & & & \textbf{Unmanned Aerial Vehicle} & \textbf{Microfilm Cartridge}
        & \textbf{Heat Exchanger} & \textbf{Automobile Brake System} \\
        \midrule
        \multirow{16}{*}{\textbf{1}}
        & \multirow{2}{*}{\textit{Mixtral-7x8B}} & \cellcolor[HTML]{EFEFEF}with    & \cellcolor[HTML]{EFEFEF}12.5±2.2 & \cellcolor[HTML]{EFEFEF}19.5±5.0 & \cellcolor[HTML]{EFEFEF}16.1±2.2 & \cellcolor[HTML]{EFEFEF}12.1±3.1 \\
        &                               & without & 15.8±4.6 & 17.7±3.3 & 16.8±1.9 & 16.0±2.3 \\
        & \multirow{2}{*}{\textit{Llama3-70B}}  & \cellcolor[HTML]{EFEFEF}with    & \cellcolor[HTML]{EFEFEF}7.6±0.8  & \cellcolor[HTML]{EFEFEF}8.3±0.5  & \cellcolor[HTML]{EFEFEF}7.9±2.0  & \cellcolor[HTML]{EFEFEF}6.1±1.1  \\
        &                              & without & 12.5±3.8 & 10.2±1.2 & 10.1±1.8 & 7.0±2.2  \\
        & \multirow{2}{*}{\textit{GPT-4-Turbo}} & \cellcolor[HTML]{EFEFEF}with    & \cellcolor[HTML]{EFEFEF}9.6±3.2  & \cellcolor[HTML]{EFEFEF}10.0±1.6 & \cellcolor[HTML]{EFEFEF}9.2±2.1  & \cellcolor[HTML]{EFEFEF}5.0±1.2  \\
        &                              & without & 13.9±4.3 & 11.0±1.5 & 12.9±3.1 & 7.5±2.7  \\
        & \multirow{2}{*}{\textit{GPT-5.2}} & \cellcolor[HTML]{EFEFEF}with    & \cellcolor[HTML]{EFEFEF}8.7±2.9  & \cellcolor[HTML]{EFEFEF}8.2±0.6  & \cellcolor[HTML]{EFEFEF}4.6±0.5  & \cellcolor[HTML]{EFEFEF}\textbf{3.0±0.0}  \\
        &                              & without & 13.9±4.1 & 8.0±0.0  & 4.9±0.7  & 3.5±0.9  \\
        & \multirow{2}{*}{\textit{DeepSeek-V3}} & \cellcolor[HTML]{EFEFEF}with    & \cellcolor[HTML]{EFEFEF}7.3±0.5  & \cellcolor[HTML]{EFEFEF}\textbf{8.0±0.0} & \cellcolor[HTML]{EFEFEF}6.6±1.0  & \cellcolor[HTML]{EFEFEF}3.9±0.5  \\
        &                              & without & 9.3±2.0 & 8.7±0.5 & 8.1±1.0 & 5.8±2.0  \\
        & \multirow{2}{*}{\textit{Qwen-3.5-Plus}} & \cellcolor[HTML]{EFEFEF}with    & \cellcolor[HTML]{EFEFEF}6.8±0.6  & \cellcolor[HTML]{EFEFEF}8.2±0.6  & \cellcolor[HTML]{EFEFEF}6.3±1.0  & \cellcolor[HTML]{EFEFEF}\textbf{3.0±0.0}  \\
        &                              & without & 9.1±2.0 & 8.6±0.9  & 6.9±1.1 & 8.1±1.3  \\
        & \multirow{2}{*}{\textit{Claude-Sonnet-3.5}} & \cellcolor[HTML]{EFEFEF}with    & \cellcolor[HTML]{EFEFEF}6.6±0.7 & \cellcolor[HTML]{EFEFEF}\textbf{8.0±0.0} & \cellcolor[HTML]{EFEFEF}4.8±0.6 & \cellcolor[HTML]{EFEFEF}4.4±0.9 \\
        &                                   & without & 11.9±3.4 & 8.1±0.3  & 5.8±1.1  & 5.9±1.4  \\
        & \multirow{2}{*}{\textit{Claude-Sonnet-4.6}} & \cellcolor[HTML]{EFEFEF}with    & \cellcolor[HTML]{EFEFEF}7.6±2.5  & \cellcolor[HTML]{EFEFEF}\textbf{8.0±0.0} & \cellcolor[HTML]{EFEFEF}5.3±1.0  & \cellcolor[HTML]{EFEFEF}3.6±0.9 \\
        &                                   & without & 15.7±7.7 & 8.3±0.5  & 5.2±0.6  & 4.1±1.4  \\
        \midrule
        \multirow{16}{*}{\textbf{5}}
        & \multirow{2}{*}{\textit{Mixtral-7x8B}} & \cellcolor[HTML]{EFEFEF}with    & \cellcolor[HTML]{EFEFEF}10.6±1.4 & \cellcolor[HTML]{EFEFEF}13.1±2.7 & \cellcolor[HTML]{EFEFEF}13.0±2.1 & \cellcolor[HTML]{EFEFEF}9.7±2.5  \\
        &                               & without & 12.3±2.6 & 15.0±2.2 & 15.4±2.3 & 13.8±1.7 \\
        & \multirow{2}{*}{\textit{Llama3-70B}}  & \cellcolor[HTML]{EFEFEF}with    & \cellcolor[HTML]{EFEFEF}6.7±0.5  & \cellcolor[HTML]{EFEFEF}8.1±0.3  & \cellcolor[HTML]{EFEFEF}5.5±0.9  & \cellcolor[HTML]{EFEFEF}5.4±1.1  \\
        &                              & without & 8.4±1.7  & 9.3±0.5  & 8.7±1.2  & 5.4±1.4  \\
        & \multirow{2}{*}{\textit{GPT-4-Turbo}} & \cellcolor[HTML]{EFEFEF}with    & \cellcolor[HTML]{EFEFEF}6.8±0.6  & \cellcolor[HTML]{EFEFEF}8.4±0.5  & \cellcolor[HTML]{EFEFEF}6.6±1.3  & \cellcolor[HTML]{EFEFEF}4.6±0.9  \\
        &                              & without & 8.9±0.8  & 9.6±0.8  & 10.1±2.7 & 5.7±1.8  \\
        & \multirow{2}{*}{\textit{GPT-5.2}} & \cellcolor[HTML]{EFEFEF}with    & \cellcolor[HTML]{EFEFEF}\textbf{6.0±0.0} & \cellcolor[HTML]{EFEFEF}\textbf{8.0±0.0} & \cellcolor[HTML]{EFEFEF}\textbf{4.0±0.0} & \cellcolor[HTML]{EFEFEF}\textbf{3.0±0.0} \\
        &                              & without & 8.4±1.7  & \textbf{8.0±0.0}  & 4.6±0.7  & \textbf{3.0±0.0}  \\
        & \multirow{2}{*}{\textit{DeepSeek-V3}} & \cellcolor[HTML]{EFEFEF}with    & \cellcolor[HTML]{EFEFEF}7.0±0.0  & \cellcolor[HTML]{EFEFEF}\textbf{8.0±0.0} & \cellcolor[HTML]{EFEFEF}5.5±0.5  & \cellcolor[HTML]{EFEFEF}3.3±0.5  \\
        &                              & without & 7.2±0.4  & 8.7±0.5  & 7.9±1.0  & 4.6±1.4  \\
        & \multirow{2}{*}{\textit{Qwen-3.5-Plus}} & \cellcolor[HTML]{EFEFEF}with    & \cellcolor[HTML]{EFEFEF}6.1±0.3  & \cellcolor[HTML]{EFEFEF}\textbf{8.0±0.0} & \cellcolor[HTML]{EFEFEF}4.7±0.9  & \cellcolor[HTML]{EFEFEF}\textbf{3.0±0.0}  \\
        &                              & without & 7.5±0.7  & \textbf{8.0±0.0}  & 5.5±0.8  & 6.1±0.7  \\
        & \multirow{2}{*}{\textit{Claude-Sonnet-3.5}} & \cellcolor[HTML]{EFEFEF}with    & \cellcolor[HTML]{EFEFEF}\textbf{6.1±0.3} & \cellcolor[HTML]{EFEFEF}\textbf{8.0±0.0} & \cellcolor[HTML]{EFEFEF}\textbf{4.0±0.4} & \cellcolor[HTML]{EFEFEF}\textbf{3.0±0.0} \\
        &                                   & without & 7.5±0.9  & \textbf{8.0±0.0} & 4.9±0.5  & 4.0±1.0  \\
        & \multirow{2}{*}{\textit{Claude-Sonnet-4.6}} & \cellcolor[HTML]{EFEFEF}with    & \cellcolor[HTML]{EFEFEF}\textbf{6.0±0.0} & \cellcolor[HTML]{EFEFEF}\textbf{8.0±0.0} & \cellcolor[HTML]{EFEFEF}4.6±0.7  & \cellcolor[HTML]{EFEFEF}\textbf{3.0±0.0} \\
        &                                   & without & 7.0±1.2  & \textbf{8.0±0.0}  & 4.4±0.8  & 3.2±0.4  \\
        \midrule
        \multirow{16}{*}{\textbf{20}}
        & \multirow{2}{*}{\textit{Mixtral-7x8B}} & \cellcolor[HTML]{EFEFEF}with    & \cellcolor[HTML]{EFEFEF}10.1±1.2 & \cellcolor[HTML]{EFEFEF}11.2±2.3 & \cellcolor[HTML]{EFEFEF}10.7±1.2 & \cellcolor[HTML]{EFEFEF}8.8±2.0  \\
        &                               & without & 11.3±1.7 & 13.0±1.2 & 14.0±1.3 & 13.1±1.7 \\
        & \multirow{2}{*}{\textit{Llama3-70B}}  & \cellcolor[HTML]{EFEFEF}with    & \cellcolor[HTML]{EFEFEF}6.7±0.5  & \cellcolor[HTML]{EFEFEF}\textbf{8.0±0.0} & \cellcolor[HTML]{EFEFEF}4.7±0.5  & \cellcolor[HTML]{EFEFEF}5.2±1.2  \\
        &                              & without & 7.4±0.8  & 8.8±0.6  & 7.7±1.3  & 4.6±0.9  \\
        & \multirow{2}{*}{\textit{GPT-4-Turbo}} & \cellcolor[HTML]{EFEFEF}with    & \cellcolor[HTML]{EFEFEF}6.1±0.3  & \cellcolor[HTML]{EFEFEF}8.2±0.4  & \cellcolor[HTML]{EFEFEF}5.1±0.8  & \cellcolor[HTML]{EFEFEF}4.1±0.7  \\
        &                              & without & 7.3±0.6  & 9.1±0.3  & 7.7±1.8  & 5.0±1.5  \\
        & \multirow{2}{*}{\textit{GPT-5.2}} & \cellcolor[HTML]{EFEFEF}with    & \cellcolor[HTML]{EFEFEF}\textbf{6.0±0.0} & \cellcolor[HTML]{EFEFEF}\textbf{8.0±0.0} & \cellcolor[HTML]{EFEFEF}3.9±0.3  & \cellcolor[HTML]{EFEFEF}\textbf{3.0±0.0} \\
        &                              & without & 6.7±0.9  & \textbf{8.0±0.0}  & 4.3±0.5  & \textbf{3.0±0.0}  \\
        & \multirow{2}{*}{\textit{DeepSeek-V3}} & \cellcolor[HTML]{EFEFEF}with    & \cellcolor[HTML]{EFEFEF}7.0±0.0  & \cellcolor[HTML]{EFEFEF}\textbf{8.0±0.0} & \cellcolor[HTML]{EFEFEF}5.4±0.5  & \cellcolor[HTML]{EFEFEF}3.1±0.3  \\
        &                              & without & 7.0±0.0  & 8.7±0.5  & 7.8±1.1  & 4.0±1.3  \\
        & \multirow{2}{*}{\textit{Qwen-3.5-Plus}} & \cellcolor[HTML]{EFEFEF}with    & \cellcolor[HTML]{EFEFEF}\textbf{6.0±0.0} & \cellcolor[HTML]{EFEFEF}\textbf{8.0±0.0} & \cellcolor[HTML]{EFEFEF}4.6±0.9  & \cellcolor[HTML]{EFEFEF}\textbf{3.0±0.0}  \\
        &                              & without & 7.0±0.0  & \textbf{8.0±0.0}  & 5.3±0.6  & 5.4±1.1  \\
        & \multirow{2}{*}{\textit{Claude-Sonnet-3.5}} & \cellcolor[HTML]{EFEFEF}with    & \cellcolor[HTML]{EFEFEF}\textbf{6.0±0.0} & \cellcolor[HTML]{EFEFEF}\textbf{8.0±0.0} & \cellcolor[HTML]{EFEFEF}\textbf{3.6±0.5} & \cellcolor[HTML]{EFEFEF}\textbf{3.0±0.0} \\
        &                                   & without & 6.4±0.7  & \textbf{8.0±0.0} & 4.1±0.3  & 3.4±0.7  \\
        & \multirow{2}{*}{\textit{Claude-Sonnet-4.6}} & \cellcolor[HTML]{EFEFEF}with    & \cellcolor[HTML]{EFEFEF}\textbf{6.0±0.0} & \cellcolor[HTML]{EFEFEF}\textbf{8.0±0.0} & \cellcolor[HTML]{EFEFEF}3.9±0.5  & \cellcolor[HTML]{EFEFEF}\textbf{3.0±0.0} \\
        &                                   & without & 6.5±1.2  & \textbf{8.0±0.0}  & 3.8±0.7  & \textbf{3.0±0.0}  \\
        \bottomrule
    \end{tabular}
    \label{tab:table5}
\end{table}

\end{document}